\documentclass{ws-ijmpd}
\usepackage[super,compress]{cite}
\usepackage{array}
\usepackage[dvips,breaklinks]{hyperref}
\hypersetup{colorlinks,urlcolor=black,citecolor=black,linkcolor=black,filecolor=black}
\usepackage{url}
\usepackage{color}

\begin{document}

\markboth{C. Negrelli, L. Kraiselburd, S. J. Landau, E. Garc\'ia-Berro}
{Spatial variation of fundamental constants: testing models with thermonuclear supernovae}

%
\catchline{}{}{}{}{}
%

\title{SPATIAL VARIATION OF FUNDAMENTAL CONSTANTS: TESTING MODELS WITH THERMONUCLEAR SUPERNOVAE}

\author{C. NEGRELLI, L. KRAISELBURD}

\address{Grupo de Astrof\'{\i}sica, Relatividad y Cosmolog\'{\i}a,\\ 
        Facultad de Ciencias Astron\'{o}micas y Geof\'{\i}sicas,\\ 
        Universidad Nacional de La Plata,\\
        Paseo del Bosque S/N 1900, La Plata, Argentina. \\
        CONICET, Godoy Cruz 2290, 1425 Ciudad Autónoma de Buenos Aires,\\
        Argentina.\\ 
cnegrelli@fcaglp.unlp.edu.ar, lkrai@fcaglp.unlp.edu.ar}
        
\author{S. LANDAU}

\address{Departamento de F\'isica and IFIBA, Facultad de Ciencias Exactas y Naturales,\\
  Universidad de Buenos Aires,\\
  Ciudad Universitaria - Pab. I, Buenos Aires 1428, Argentina.\\
 CONICET, Godoy Cruz 2290, 1425 Ciudad Autónoma de Buenos Aires, Argentina. 
slandau@df.uba.ar}        

\author{E. GARC\'IA-BERRO$^\dagger$\footnote{deceased}}

\address{Departament de F\'isica, Universitat Politecnica de Catalunya,\\
 c/Esteve Terrades 5, 08860 Castelldefels.\\
Institute for Space Studies of Catalonia,\\
 c/Gran Capita 2–4, Edif. Nexus 201, 08034 Barcelona.\\
Spain}     

\maketitle

\begin{history}
\received{Day Month Year}
\revised{Day Month Year}
\end{history}

\begin{abstract}
Since Dirac  stated his {\it  Large Number Hypothesis}  the space-time
variation  of fundamental  constants  has been  an  active subject  of
research.  Here we analyze the possible  spatial variation of
two fundamental constants: the fine structure constant $\alpha$ and the
speed of light $c$. We  study the  effects of such  variations on  the luminosity
distance and on the peak luminosity  of Type Ia supernovae (SNe Ia).  For this,
we consider  the change  of each  fundamental constant  separately and
discuss  a  dipole model  for  its  variation.  Elaborating  upon  our
previous  work,  we  take  into  account the  variation  of  the  peak
luminosity of Type Ia supernovae  resulting from the variation of each
of  these fundamental  constants.   Furthermore, we  also include  the
change  of the  energy release  during  the explosion,  which was  not
studied before  in the  literature. We perform a statistical analysis to compare the predictions of the dipole model for $\alpha$ and $c$ variation with the Union2.1 and JLA compilations of SNe Ia. For this, we also allow the nuisance parameters of the distance estimator $\mu_0$ and the cosmological density matter $\Omega_{\rm m}$ to vary.   As  a result  of our  analysis we
obtain a first estimate of the possible spatial variation of the speed
of light $c$. On the other hand, we find that
there   is    no   significant   difference   between    the   several
phenomenological  models studied  here and  the standard  cosmological
model, in  which fundamental constants do  not vary at all.   Thus, we
conclude that  the actual set of  data of Type Ia  supernovae does not
allow to  verify the hypothetical spatial variation  of fundamental
constants.
\end{abstract}


\ccode{PACS numbers: 06.20.Jr, 98.80.-k, 98.80.Es, 97.60.Bw}

\section{Introduction}
\label{sec:intro}

The Standard Model of Cosmology  describes, in good agreement with the
observations, the  evolution of the  Universe.  The Standard  Model is
based  on  the  assumption  that fundamental  constants  -- the fine structure  constant $\alpha$,
the  proton-to-electron  mass ratio  $\mu$ and the
gravitational constant  $G$ among others --  remain truly  constant
throughout space and time.   However, there  are other  competing theories
that do  not rely in  this, otherwise reasonable,  hypothesis. Indeed,
theories that predict  a variation of fundamental  constants have been
developed over the years. They can be divided into two categories. The
first  one of  these families  are grand-unification  theories. Within
this theoretical  framework fundamental  constants are  slowly varying
functions of  low-mass dynamical scalar  fields -- see,  for instance,
Refs.~\refcite{Pablo,Uzan,Yuri},  and  references
therein.  The second family  of formulations corresponds to low-energy
effective  theories.  These  are phenomenological  models specifically
proposed to study a potential  variation of fundamental constants.  In
these  theories  the ``parameter''  whose  variation  is going  to  be
studied   is   replaced  by   a   scalar   field.  Furthermore,   each
phenomenological model assumes that only a fundamental coupling varies
at   a   time,   either   the   fine   structure   constant   $\alpha$
\cite{bekenstein82,bekenstein2002,bsm02},  the  speed  of  light  $c$
\cite{moffat2,albrecht,barrow} or the electron mass \cite{BM05}. 

Different versions  of the theories mentioned  above predict different
variations for the fundamental constants. Thus, if these theories turn
out to  be correct, the  fundamental constants are expected  to depend
weakly on  time, or  vary on  different space  lengths. In  essence, this  is equivalent to  say that  searching for
possible  variations  in the  fundamental  constants  of nature  would
eventually allow us  to test whether the laws of  physics are the same
everywhere at any time. Therefore, it is important to analyze the observational effects that may arise  from changes of the fundamental constants, and to develop new methods to constrain these hypothetical  variations in order to check the validity of such theories. 

Numerous  experiments and  observations  have  attempted to  establish
whether  the   fundamental couplings  are indeed  constant.
The  experimental studies  can be  grouped in  two broad  classes. The
first of them comes from purely  local methods, whereas the second one
is  based in  the observation  of astronomical  phenomena. The  former
category  includes geophysical  methods  such as  the natural  nuclear
reactor that  operated about $1.8  \times 10^9$~yr ago in  Oklo, Gabon
\cite{dam3,Petrov06,Gould06},  the  analysis  of  natural  long-lived
$\beta$ decays in old  minerals and meteorites \cite{dyson, sisterna,
Olive04b},  and laboratory  measurements,  that  include for  instance
detailed comparisons of several  atomic clocks with different isotopes
\cite{pres,Peik04,rosenband08}. The latter family of methods is based
mainly, but  not only, on  the analysis of spectra  from high-redshift
quasar absorption systems \cite{bahc,  lev, murphy1, murphy2, Webb99,
webb2}.  Besides,  further constraints on a  hypothetical variation of
$\alpha$   can   be   obtained   by  comparing   X-ray   results   and
Sunyaev-Zeldovich      measurements       in      galaxy      clusters
\cite{Galli13,Holanda16,Holanda16b,Martino16,Martino16b}.   Moreover,
the variation of $\alpha$ in the  early universe can be constrained as
well  from primordial  nucleosynthesis \cite{bergstrom,mosquera}  and
from  the  Cosmic  Microwave  Background  (CMB)  fluctuation  spectrum
\cite{bat,avelino,Planck2015,obryan15}.     We recommend the interested reader a careful reading of    the    reviews   of
Refs.~\refcite{Uzan} and ~\refcite{Yuri} for  extensive discussions of  the many
observational techniques.

Evidence for a dipole spatial variation of $\alpha$ was obtained using
the  combined observations  of  distant quasars  using the  KECK/HIRES
\cite{Murphy03b}   and  VLT/UVES   \cite{Webb11,King12}  telescopes.
However,  subsequent  analyses  \cite{whit}  showed  that  long-range
wavelength distortions can mimic the  effect of the reported variation
of $\alpha$. On the other  hand, Ref.~\refcite{PM16} performed an independent
analysis using the observed data  set, together with other 
observational results and found  that the observations were consistent
with  a dipole  variation of  $\alpha$.  Finally, recent  work on  two
Zn{\sc II} and three Cr{\sc III}  transitions has provided us with the
first method to test a variation of $\alpha$ that is not influenced by
long-range  distortions \cite{Murphy16,Murphy17}.  Even though  these
latest  results  showed no  convincing  evidence  for a  variation  of
$\alpha$, Ref.~\refcite{Murphy17}  concluded that their quasar  sample is too
small to rule  out the dipole model. Otherwise,  the spatial variation
of   $\alpha$    has   been    analyzed   using    different   methods
\cite{Galli13,Martino16,Martino16b,obryan15,Iorio11,Li15}.         In
particular,   in   addition   to   the   analyses   mentioned   above,
Ref.~\refcite{obryan15} obtained  constraints from  the CMB  radiation, while
Ref.~\refcite{Iorio11}  studied  the effect  of  the  CMB modulation  on  the
orbital  motion  of the  major  bodies  of  the Solar  System.   Also,
Ref.~\refcite{Li15} studied a Finslerian Universe where both $\alpha$ and the
luminosity distance  of Type Ia  supernovae require  a dipole
variation.

On the  other hand, there is  some evidence of a  dipole anisotropy in
other  cosmological  observables.   For  instance,  Refs.~\refcite{YWC14}  and ~\refcite{Mariano} proposed a dipole model  for the deviation of distance
modulus  of  SNe~Ia  with  respect   to  the  standard  value  of  the
$\Lambda_{\rm CDM}$ model, and  performed a statistical analysis using
the Union~2 and Union~2.1 samples.  Their results show that the dipole
that best  fits the  SNe~Ia sample  has a  similar direction  than the
dipole that results from the quasar sample. Nonetheless, none of these
authors considered the effects on the luminosity distance of SNe~Ia of
a variation  of $\alpha$.  On  the other hand,  Ref.~\refcite{Mariano13} have
identified a direction  of the maximum temperature  asymmetry (MTA) of
the WMAP7 reduced map and found that the direction of the asymmetry is
consistent  with  that  found  by the  previously  mentioned  analyses
\cite{YWC14,Mariano}.  However, Ref.~\refcite{Chang15}  concluded that using
two  different  methods,  i.e.,  dipole-fitting    and  hemisphere
comparison, the preferred directions coming from the Union~2 data
are  approximately opposite.  We  note, however,  that  later the  two
methods have  been critically reviewed, and  it was found that  the dipole-fitting
method is  statistically significant while  the hemisphere comparison method  is strongly
biased by the  distribution of data points in  the sky \cite{Lin16b}.
Furthermore,  Ref.~\refcite{Lin16}   preformed  a  statistical   analysis  to
constrain  the amplitude  and direction  of anisotropy  of the  SNe~Ia
using  a new  data set,  the JLA  compilation \cite{Betoule14}.  They
applied a  Markov Chain Monte Carlo  method and they derived  a dipole
direction   that    is   not   consistent   with    the   results   of
Ref.~\refcite{Mariano}. Nevertheless,  they have obtained  consistent results
when  applying  the same  method  to  the Union~2.1  compilation.   In
summary, the anisotropy  derived from SNe~Ia strongly  depends on both
the employed data sets, and on the methods used to analyze the data.

Additionally,  there is  some  observational evidence  that could  be
interpreted  as a  hint  for deviations  from large-scale  statistical
isotropy such as the large-scale
alignment in the QSO  optical polarization data \cite{quasarspol} and the  alignment of low  multipoles in  the CMB
angular  power  spectrum  \cite{anomaliesCMB}. These observations  have intensified  the
interest in the spatial variation of fundamental constants.

In a previous work \cite{krai}, we  studied the effects of a possible
spatial variation of $\alpha$ on the luminosity distance of SNe~Ia. In
that work we  included the variation of the  peak luminosity resulting
from  the variation  of $\alpha$,  which was  not analyzed  before. In
particular,   we    used   the   previous   analysis    performed   by
Ref.~\refcite{ChibaKohri}, who considered the  dependence of the mean opacity
of the expanding  photosphere of SNe~Ia on the value  of $\alpha$, and
in addition  we evaluated  the effects  of a  varying $\alpha$  on the
precise value of the Chandrasekhar  limit. Both physical effects change
the  luminosity distance  of  SNe~Ia.  In  this work  we  go one  step
further  and consider  the  variation of  two different  fundamental
constants  separately.

The authors of Ref.~\refcite{MBS02} have discussed the differences between theories where $e$ is the varying fundamental constant with respect to  those where $c$ varies. One of the most important differences is that in the former case, the Weak Equivalence Principle is violated while in the latter it is not. On the other hand, models where the variation in $c$ is spatial were analyzed in Ref.~\refcite{Magueijo01}.
Therefore, we consider first  a   variation  of the  fine
structure constant  $\alpha$, and then a  varying speed of light  $c$.
 It is important to stress that the variation of  $\alpha$ considered in this paper  is different from
that considered  in our  previous work  \cite{krai}.  In Ref.~\refcite{krai}
 we considered the  spatial variation of $\alpha$, while
keeping all  other fundamental constants  fixed. In contrast,  in this
paper we  analyze the variation  of $\alpha$ through the  variation of
the  electron charge  $e$.  As  a  consequence the  dependence of  the
relevant quantities for the peak  luminosity with $\alpha$, namely the
opacity  of the  expanding  photosphere, and  the Chandrasekhar  mass,
differ from those  obtained in our previous  analysis. Furthermore, we
investigate  the effect  of a  possible variation  of the  fundamental
constants on  the energy  release during  the supernova  explosion. In
addition, we propose a dipole model  for the spatial variation of each
fundamental constant considered  in this paper. Finally,  we perform a
statistical  analysis using  the distance  modulus of  SNe~Ia obtained
from  the Union~2.1  \cite{Suzuki12}  and  the JLA  \cite{Betoule14}
compilations to check if the  models are compatible with observations taking into account the emerging estimates of the nuisance parameters and of the cosmological density matter.  The reason for this is that  the standardization of the SNe Ia depends on the theoretical model used for the distance modulus, in the sense that it influences the nuisance parameters determination which accompany the stretch, color and host-mass corrections. 

Our paper is organized  as follows.  In Sect.~\ref{sec:dependencia} we
analyze the dependence of the peak  luminosity of SNe~Ia peak with the
variation of fundamental constants and  how a possible change modifies
the the distance modulus.  It follows Sect.~\ref{sec:dipolo}, where we
present  the  dipole  model. Then,  in  Sect.~\ref{sec:resultados}  we
present   our  results.    Lastly,  in   Sect.~\ref{sec:discusion}  we
summarize our main findings and we draw our conclusions.

\section{The Distance Modulus of Thermonuclear Supernovae}
\label{sec:dependencia}

SNe~Ia are among the most energetic and
interesting  phenomena   in  our  universe.   Owing   to  their  large
luminosities, they can be observed  up to very high  redshifts. 
Moreover,  a sizeable  number of  them are nowadays routinely detected  by dedicated
surveys. All this makes them suitable astronomical objects to test the
possible  spatial variation  of  fundamental constants. But  not  only that,  the
spectra and  light  curves of normal SNe~Ia are very homogeneous. This arises because the light
curve  of a  SNe~Ia can  be  understood in  terms of  the capture  and
thermalization  of  the  products  of  radioactive  disintegration  of
$^{56}$Ni  and  $^{56}$Co.   SNe~Ia  reach their  peak  luminosity   in  approximately  20  days  after  explosion
\cite{riess}.  Moreover, it is observationally  found that there is a
tight  correlation between  their peak  bolometric magnitudes  and the
decline rates of  their light curves.  All these  features make SNe~Ia
one of the best standard candles known today.

An other important property of SNe~Ia is that they are detected in all
types of galaxies.  It follows from  this, and from the homogeneity of
the observed characteristics  of SNe~Ia, that normal SNe~Ia  share the same
explosion  mechanism. The  details  of this  mechanism  are still  the
subject of active  research. However, it is well  established that the
homogeneity of the light curve is essentially due to the narrow spread
of  nickel  masses  (${\rm{M}}_{\rm  {Ni}}\sim  0.6 \ \rm{M_{\odot}}$)
produced in the  explosion of a carbon-oxygen white dwarf  with a mass
close  to the  Chandrasekhar limit.   Consequently, the  observational
properties of SNe~Ia are primarily  determined by the precise value of
the Chandrasekhar limiting mass.

\subsection{The dependence of the intrinsic properties of SNe~Ia on
  $\alpha$ and $c$}

The  maximum  mass of  a  stable  white dwarf  star  is  given by  the
Chandrasekhar  limit.   Above this  mass,  the  pressure of  degenerate
electrons cannot  balance the gravitational  force.  The value  of the
Chandrasekhar  mass  is  $\simeq 1.44\,{\rm{M_{\odot}}}$  and  can  be
expressed as:
\begin{equation}
\label{chandra}
  M_{\rm Ch}=\frac{w_3^0 \sqrt{3 \pi}}{2} \left(\frac{\hbar c}{G} \right)^{3/2} \frac{1}{(\mu_e m_{\mathrm{H}})^2},
\end{equation}
where  $\mu_e$   is  the   average  molecular  weight   per  electron,
$m_{\mathrm{H}}$  is the  mass  of  the hydrogen  atom,  $w_3^0$ is  a
constant  and $G$  is the  gravitational constant. In  our  analysis  we  will assume  that  the
variation of $\alpha$  follows from a variation in $e$. Note that the
Chandrasekhar    limiting   mass    depends    only    on   $c$. Consequently,  if this constant  varies, the mass  limit will
change accordingly,  and so will  do the  mass of nickel synthesized in the explosion. This,  in turn,
will eventually lead  to a different peak luminosity  of SNe~Ia, and
will  ultimately  affect the  determination  of  distances to  distant
supernovae.   In particular,  a small  variation of  $c$ results  in a
variation of the Chandrasekhar mass:
\begin{equation}
\label{masac}
  \frac{\delta M_{\mathrm{Ch}}}{M_{\mathrm{Ch}}}=\frac{3}{2}\frac{\delta c }{c }.
\end{equation}
At this  point we would like  to emphasize that the  treatment adopted
here differs from that used in our previous work in which we assumed a
varying $\alpha$. When this is the  case, there is a dependence of the
Chandrasekhar mass  with $\alpha$.  However, in  the present  work the
variation  of $\alpha$  arises from  a hypothetical  variation of  the
electron mass. Therefore, there is  no dependence of the Chandrasekhar
mass with $\alpha$.

One of the main improvements that  incorporates this work is the study
the energy  released during the  explosion.  The energy released  in a
SNe~Ia outburst comes from the  difference of nuclear binding energies
of  nickel and  cobalt. The  the leading  contribution is  the Coulomb
term:
\begin{equation}
  E=\frac{3}{5} \frac{e^2}{r_0}\frac{Z^2}{A^{1/3}}, 
\end{equation}
where $Z$  is the atomic  number, $A$ is  the number of  nucleons, and
$r_0$ an empirical constant. Therefore:
\begin{equation}
  \frac{\delta E}{E}=\frac{\delta \alpha }{\alpha }.
  \label{ealpha}
\end{equation}
Thus, the energy released depends only on the value of $\alpha$, and
 not  on $c$.

The peak luminosity of SNe~Ia not  only depends on the energy released
during  the  explosion, but  also  on  the  opacity of  the  expanding
photosphere. Actually,  the emitted photons do  not escape immediately
because   the  material   ejected   in  the   outburst  is   optically
thick. At early times, the opacity is dominated by the line of opacity $\kappa_i$ rather than the electron scattering contribution $\kappa_e$.  It should be noted that $\kappa_i \sim \alpha$ while $\kappa_e \sim \alpha^2$. However, since our analysis focuses on the long-term variations of the observed luminosities due to a hypothetical variation of $\alpha$, we adopt the treatment of Ref.~\refcite{ChibaKohri}, which is a simplified model of the explosion. We note, nevertheless, that a more detailed analysis can be found in  Ref.~\refcite{Karp77}. Within this 
approximation:
\begin{equation}
  \kappa \sim \kappa_e = \frac{n_e}{\rho} \sigma _{\rm {Th}},
\end{equation}

where $n_e$  is the  number of  electrons, $\rho$  is the  density and
$\sigma$ is the Thomson cross section:
\begin{equation}
  \sigma_{\rm {Th}}=\frac{8 \pi}{3} \left(\frac{e^2}{m c^2}\right)^2,
  \label{thom}
\end{equation}
being $m$ the electron mass. This expression can be rewritten in terms
of $\alpha$:
\begin{equation}
  \sigma_{\rm {Th}}=\frac{8 \pi}{3} \left(\frac{\alpha \hbar}{m c}\right)^2,
\end{equation}
and therefore:
\begin{equation}
\label{opacidadalfa}
  \frac{\delta \kappa }{\kappa }=2\frac{\delta \alpha }{\alpha }.
\end{equation}
Also, from Eq.~(\ref{thom}) we find that:
\begin{equation}
\label{opacidadc}
  \frac{\delta \kappa }{\kappa }=-4\frac{\delta c }{c }.
\end{equation}
Thus,  the opacity  depends on  $\alpha$ and  $c$.

In  Table~\ref{tab:resumen},  we  show   a  summary  of  the  previous
analysis.   There  we  list  the different  dependencies  of  the  peak
bolometric  magnitude on  fundamental constants.  In the  following we
calculate  how  the  peak  bolometric magnitude  scales on  $\alpha$  and  $c$,  taking  into account  all  the  dependencies  just
described.

\begin{table}[ph]
\tbl{Variation  of   the  physical  quantities  involved   in  the
  supernovae explosion due to the variation of fundamental constants.}
  {\begin{tabular}{@{}cccc@{}} \toprule
Varying constant    & {\large  $\frac{\delta M_{\mathrm{Ch}}}{M_{\mathrm{Ch}}}$} & {\large  $\frac{\delta E}{E}$} & {\large $\frac{\delta \kappa}{\kappa}$}\\   \colrule
$\alpha$            & $0$ &  $\displaystyle\frac{\delta \alpha}{\alpha}$ & $2\displaystyle\frac{\delta \alpha}{\alpha}$ \\
$c$                 & $\displaystyle\frac{3}{2} \displaystyle\frac{\delta c}{c}$ & $0$ & $-4\displaystyle\frac{\delta c}{c}$\\\botrule
\end{tabular} \label{tab:resumen}}
\end{table}

\subsection{The peak luminosity and its dependence on $\alpha$ and $c$}

The relation between the fundamental constants and the peak bolometric magnitude of SNe~Ia can be  obtained using simple analytical arguments. We
follow  closely the procedure  of Ref.~\refcite{krai} which relies on the analysis of  Ref.~\refcite{ChibaKohri}, but this time taking into
account all the dependencies on $\alpha$ and $c$. The peak luminosity of the optical light curve is essentially proportional to the energy deposition rate of the $^{56}$Ni$\rightarrow^{56}$Co$\rightarrow^{56}$Fe decay chain inside the photosphere of the exploding star at the time $t_{\rm peak}$, which is the time where the diffusion and expansion timescales are similar ($t_{\mathrm{peak}}    \sim t_{\mathrm{exp}}   \sim  t_{\mathrm{dif}}$). It is important to note that the energy deposition rate  depends mainly on  $t_{\rm peak}$. Moreover, the $\gamma$-ray  deposition   function   can be developed in a power series:
\begin{equation}
\frac{\delta G}{G}=\frac{1.6}{1.6+3.6} \frac{\delta \tau }{\tau },
\end{equation}
where $\tau$ is the optical depth, for which we adopt  $\tau_{\mathrm{peak}}\sim 3.6$.
Then,  the change of the
peak luminosity as a function of $t_{\rm peak}$ and $\tau_{\mathrm{peak}}$ is given by: 
\begin{equation}
\label{lum}
\frac{\delta L_{\rm peak}}{L_{\rm peak}}=-\frac{\delta t_{\mathrm{peak}}}{t_{\mathrm{peak}}}+\eta\frac{1.6}{1.6+3.6} \frac{\delta \tau_{\mathrm{peak}}}{\tau_{\mathrm{peak}}},
\end{equation}
being
\begin{equation}
\eta=1+4G(t_{\mathrm{peak}})-10.5G(t_{\mathrm{peak}})^2+6G(t_{\mathrm{peak}})^3.
\end{equation}

In what  follows, we  will analyze the dependence of $t_{\rm
peak}$ and  $\tau_{\rm peak}$ with each fundamental constant considered in this paper.

\subsubsection{Model A: varying $\alpha$}

A   variation  of the  fine structure  constant would  result in  a
change  of  the opacity  and of  the  energy  of the  explosion. Accordingly,  $t_{\rm peak}$ will be also modified. This happens because
\begin{equation}
\label{tpeak}
t_{\rm peak}=\left(\frac{3\kappa}{4\sqrt{2}\pi c}\right)^{1/2}
            \left(\frac{M_{\mathrm{Ch}}^3}{E}\right)^{1/4},
\end{equation}
and
\begin{equation}
  \label{tau}
  \tau_{\mathrm{peak}}=\frac{\sqrt{2}c}{2}\left(\frac{M_{\mathrm{Ch}}}{E}\right)^{1/2},
\end{equation}
For additional details we refer the interested reader to  Ref.~\refcite{krai}, but we note that in  Eq.~(14) of this paper a factor $1/2$ is missing. Then,
\begin{equation}
 \label{tke}
  \frac{\delta t_{\mathrm{peak}}}{ t_{\mathrm{peak}}}=\frac{1}{2}\frac{\delta \kappa }{\kappa }-\frac{1}{4} \frac{\delta E}{E},
\end{equation}
and
\begin{equation}
\label{taue}
  \frac{\delta \tau_{\mathrm{peak}} }{\tau_{\mathrm{peak}} }=-\frac{1}{2}\frac{\delta E}{E}.
\end{equation}
Combining   Eqs.~(\ref{ealpha}),  (\ref{opacidadalfa}),   (\ref{lum}),
(\ref{tke}) and (\ref{taue}),  the variation  of the peak
luminosity in  terms of the  variation of the fine  structure constant
$\alpha$ can be written as:
\begin{equation}
  \frac{\delta L_{\mathrm{peak}}}{L_{\mathrm{peak}}}\simeq -0.8269 \frac{\delta \alpha }{\alpha}.
\end{equation}

\begin{figure}[pb]
\centering
    \includegraphics[width=0.6\textwidth,trim=0cm 4.0cm 0.5cm 3.6cm,clip=true]{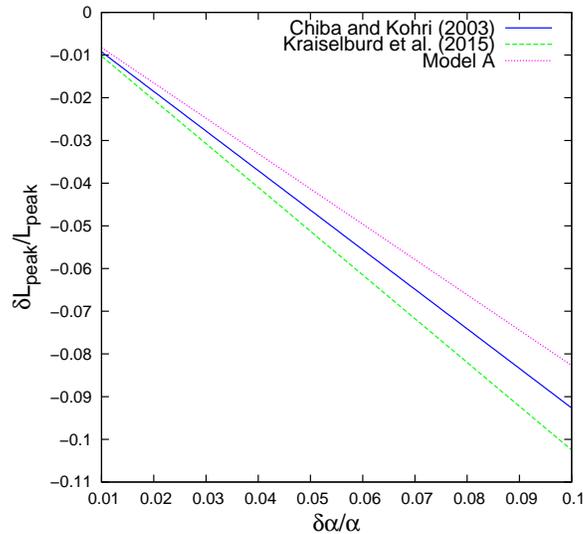}
    \vspace*{8pt}
    \caption{Peak  luminosity  of  distant  SNe~Ia as  a  function  of
      $\delta\alpha/\alpha$. Here  Model A corresponds to  the case in
      which the  changes of the  energy released during  the explosion
      and the opacity of the expanding photosphere are considered.}
\label{fig:dlda}
\end{figure}

In  Fig.~\ref{fig:dlda}   we  compare   our  results  with   those  of
Ref.~\refcite{ChibaKohri} and Ref.~\refcite{krai}. As can be seen, the dependence of
$\delta L_{\rm  peak}/L_{\rm peak}$ on  $\delta\alpha/\alpha$ obtained
by  Ref.~\refcite{krai}  has a  smaller  slope  than  the one  calculated  by
Ref.~\refcite{ChibaKohri}.   Conversely,   the  relation  between   the  peak
luminosity and the variation of $\alpha$  computed in this paper has a
slope stepper than  previous estimates.  This, clearly, is  due to the
fact that  in the present  work we do not  only take into  account the
dependence  of the  opacity on  $\alpha$.  Also,  here we  consider the
dependence of the energy released in the supernova outburst, which was
not taken into  account by Ref.~\refcite{ChibaKohri}. However,  we stress that
our results and those of Ref.~\refcite{ChibaKohri} show that a decrease of the
value of  $\alpha$ translates  into an increase  of the  luminosity of
thermonuclear supernovae.  Thus a  smaller (larger) value  of $\alpha$
makes SNe~Ia brighter (fainter).

\subsubsection{Model B: varying $c$}

According  to the  previous discussion  -- see  Table~\ref{tab:resumen} --
within  this scenario  both  the opacity  and  the Chandrasekhar  mass
vary. Using Eqs. (\ref{tpeak}) and (\ref{tau}) we get:
 \begin{equation}
 \label{tpeakc}
  \frac{\delta t_{\mathrm{peak}}}{ t_{\mathrm{peak}}}=-\frac{1}{2}\frac{\delta c}{c} + \frac{1}{2}\frac{\delta \kappa }{\kappa }+ \frac{3}{4}\frac{\delta M_{\mathrm{Ch}}}{M_{\mathrm{Ch}}},
\end{equation}
and
\begin{equation}
\label{taualfa2}
  \frac{\delta \tau_{\mathrm{peak}} }{\tau_{\mathrm{peak}} }=\frac{\delta c}{c}+\frac{1}{2}\frac{\delta M_{\mathrm{Ch}}}{M_{\mathrm{Ch}}}=\frac{7}{4}\frac{\delta c }{c }.
\end{equation}
Combining    Eqs.~(\ref{masac}),    (\ref{opacidadc}),    (\ref{lum}),
(\ref{tpeakc}) and (\ref{taualfa2}) we can obtain the variation of the
peak luminosity due to a variation on the speed of light $c$:
\begin{equation}
  \frac{\delta L_{\mathrm{peak}}}{L_{\mathrm{peak}}}\simeq 1.6442\frac{\delta c }{c}.
\end{equation}
Note that an increase of the  value of $c$ translates into an increase
of the luminosity of thermonuclear supernovae. Thus a larger (smaller)
value of $c$ would make SNe~Ia brighter (fainter).

\subsection{The distance modulus}
\label{sec:moddist}

The  distance  modulus  of  a  supernova in  the  case  of  a  varying
fundamental constant can be expressed as:
\begin{equation}
   \mu=25+5 \log{\left( \frac {d_L}{\mathrm{Mpc}}\right)}+\delta M,
\end{equation}
where $d_L$ is the standard distance
modulus  that depends on the cosmological parameters and redshift $z$,  and $\delta  M$  is  a correction  taking  into account  such
variation:
\begin{equation}
  \delta M=-2.5\frac{\delta L_{\mathrm{peak}}}{L_{\mathrm{peak}}}.
\end{equation}

For our  calculations we adopt  a flat $\Lambda_{\rm CDM}$  model with
  $\Omega_{\rm R}=0.0$ and $H_0=67.31\,{\rm   Mpc^{-1}\,  km   \,   s^{-1}}$.  The latter  is    the best fit   value    obtained   by   the    Planck   collaboration
\cite{Planck16}.  It was  derived   using  the  Cosmic  Microwave
Background  (CMB)  temperature   ($30  <  l  <   2508$),  the  low-$l$
polarization data  ($2 < l <  29$), together with the  position of the
peak       of       the       Baryon       Acoustic       Oscillations
\cite{Beutler11,BAO1,Ross15}. Besides, the matter density $\Omega_m$ is a free parameter in our analysis. 

\section{Dipole Models}
\label{sec:dipolo}

In subsequent sections we will compare the observed distance moduli of
SNe~Ia with the  theoretical predictions.  For such  comparison we use
the previously explained cosmological  model, but including a possible
variation       of       fundamental      constants       --       see
Sect.~\ref{sec:dependencia}.   To  account   for   the  variation   of
fundamental constants we adopt a dipole model.  At this point we would
like to  emphasize that even  though dipole models  are controversial,
the  recent observational  evidence is  not conclusive,  and does  not
allow to safely discard them.

In  particular, we  adopt  the following  expression  for the  spatial
variation of $\alpha$:
\begin{equation}
\frac{\delta \alpha}{\alpha}=A_{\alpha}+B_{\alpha}  \cos\theta_{\alpha},
\label{alphavar}
\end{equation}
where $\cos\theta_{\alpha} = \vec{r}  \cdot \vec{D}$, $\vec{D}$ is the
direction  of  the dipole,  $\vec{r}$  is  the  position on  the  sky,
$A_{\alpha}$ is a  constant (a monopole term) and  $B_{\alpha}$ is the
amplitude of the dipole term.
Likewise,  for the  variation of  $c$   we adopt  a similar
expression:
\begin{equation}
  \frac{\delta c}{c}=  A_c+B_c \cos \theta_c 
\end{equation}

\section{Results}
\label{sec:resultados}

We  now proceed  to compare  the predictions  of the  phenomenological
dipole  models  discussed  earlier  with the  data  of  the  Union~2.1
\cite{Suzuki12}  and JLA  \cite{Betoule14}  compilations of  SNe~Ia.
These datasets do not incorporate either 
  luminous supernovae nor fast and bright transients, Ca-rich transients or .Ia supernovae, but only ``normal'' or classical thermonuclear supernovae that
  follow closely the canonical  relationship between its intrinsic brightness and the decline rate of the light curve \cite{Phillips}. Hence the intrinsic dispersion of the corresponding
Hubble diagram is much smaller.
Specifically, using these high-quality datasets we  consider  the  values  of $A$,  $B$  and  $\vec  D$
introduced in Sect.~\ref{sec:dipolo} as free parameters, and we obtain
the best-fitting values using the observational data provided by these
compilations.

To  quantify the  agreement between  the theoretical  results and  the
observed  data we use  a
$\chi^2$  test.  Acording to Ref. \refcite{Suzuki12}, the distance modulus estimator for the Union~2.1 compilation can be expressed as:

\begin{equation}
  \mu_{\rm {o}}=m_b^*-(M_b-\widetilde{\alpha} X_1+\widetilde{\beta} C +\widetilde{\delta} P),
\end{equation}

 where $m_b^*$ corresponds to the observed peak magnitude in the $B$ band, $X_1$ and $C$ refer  to the deviation from the average light-curve shape and the mean SN Ia B−V color respectively \footnote{The light-curve parameters $X_1$ and $C$
result from the fit of a model of the SNe Ia spectral sequence to the photometric data (for details see Ref. \refcite{Suzuki12}).}, and $P = P(m_*^{\rm host} < m_*^{\rm Threshold})$ is the probability that the true mass of the host galaxy is less
than a certain  mass threshold. Besides, $M_b$, $\widetilde{\alpha}$, $\widetilde{\beta}$ and $\widetilde{\delta}$ are the nuisance parameters.

In this case the $\chi^2$  estimator  is:
\begin{equation}
\chi^2_{\mathrm{Union2.1}} = \sum_i \frac{\left(\mu_{\rm {t_i}} - 
\mu_{\rm {o_i}}\right)^2}{\sigma_{\rm {o_i}}^2},
\end{equation}
where the  subscript $i$ refers  to each observational data  point, whereas $\sigma_{\rm {o_i}}$ the total errors including systematics and sample dependent effects are taken from the covariance matrix of the  Union~2.1 compilation~\cite{Suzuki12}.

On the other hand, the distance modulus for the JLA compliation reads~\cite{Betoule14}:

\begin{equation}
  \mu_{\rm {o}}=m_b^*-(M_b + \Delta_{\rm M} -\widetilde{\alpha} X_1+\widetilde{\beta} C),
\end{equation}
where $m_b^*$, $X_1$ and $C$ where described above and $\Delta_{\rm M}$ is a nuisance parameter which is set to $0$ if the star host mass is lower than $10^{10} M_{\odot}$. Again, $M_b$, $\widetilde{\alpha}$, $\widetilde{\beta}$ are  nuisance parameters. Then, the  $\chi^2$   estimator  computed  using  the
expression of Ref.~\refcite{Betoule14} is:
\begin{equation}
\chi^2_{\mathrm{JLA}} = \left(\mu_{\rm {o}} - \mu_{\rm {t}}\right)^{\rm {T}} S^{-1} \left(\mu_{\rm {o}} - \mu_{\rm {t}}\right) ,
\end{equation}
with $S$ the covariance matrix of $\mu_{\rm {o}}$. The data of these magnitudes used for the calculation as well as the covariance matrix are taken from the JLA compilation.

%

We   calculate  the  reduced $\chi^2$  -- that  is  the value  of
$\chi^2$ divided by the number of degrees of freedom, $\nu$ -- for the
standard cosmological model  for the case in which we  do not consider
any  hypothetical variation  of  the fundamental  constants. For  the
complete data set of the  Union~2.1 compilation ($580$ data points) we
obtain $\chi^2/\nu=1.11$ while   for the JLA compilation ($ 740$ data points)  we
obtain  $\chi^2/\nu=0.96$.    As  already  noted  in   previous  works
\cite{Planck14,  Planck16}  the  agreement between  the  cosmological
parameters  obtained by  the Planck  collaboration and  those obtained
using the JLA  compilation is considerably better  than those obtained
when the Union~2.1 dataset is  employed.  However, this discrepancy is
within the $1\sigma$ error bar.

Next, we perform the statistical analyzes to compare the theoretical prediction for  the distance modulus
$\mu_{\rm {t}}$ including the possible variation of each fundamental constant ($\alpha$ or $c$) according to the
phenomenological  dipole   model described in Sect.\ref{sec:dipolo} with observational data from SNIa. For each data set we consider the following  free parameters:

\begin{enumerate}
\item{For the analyzes performed with data from the  JLA compilation: the matter density in units of the critical density $\Omega_m$,  the nuisance parameters $\widetilde{\alpha}$, $\widetilde{\beta}$, $M_b$ and $\Delta M$ and the dipole model parameters $A$, $B$, the right ascention $R.A$ and declination $\delta$ of the dipole model }
\item{For the analyzes performed with data from the Union 2.1 compilation: the matter density in units of the critical density $\Omega_m$,  the nuisance parameters $\widetilde{\alpha}$, $\widetilde{\beta}$, $M_b$ and $\widetilde{\delta}$ and the dipole model parameters $A$, $B$, the right ascention $R.A$ and declination $\delta$ of the dipole model }
\end{enumerate}

\begin{table}[ph]
\tbl{Results for the parameters of the dipole model for the spatial variation of $\alpha$, the SNe Ia nuisance parameters and $\Omega_{\rm m}$ obtained from the statistical analysis with the 1$\sigma$ error. $\Delta$ refers to $\Delta_{\rm M}$ for the analysis performed with the JLA data and to $\widetilde{\delta}$ for the Union2.1 data.The size of the JLA dataset is 740 SNe Ia. The size of the Union2.1 dataset is 580 SNe Ia. For the standard model the value of the goodness of fit is $\chi^2_{\nu}$=0.96 for JLA and $\chi^2_{\nu}$=1.11 for Union2.1.}  
{\begin{tabular}{@{}cccccccccccc@{}}\toprule
 Dataset & $A_{\alpha}$  & $B_{\alpha}$ &  R.A. & $\delta$ & $\widetilde{\alpha}$ &  $\widetilde{\beta}$ & $\Delta$ & ${\rm M}_{\rm b}$& $\Omega_{\rm m}$& $\chi^2_{\nu}$ \\  
 & ($\times 10^{-2}$) & ($\times 10^{-2}$) & (h) & ($^{\circ}$)  & ($\times 10^{-1}$) & & ($\times 10^{-2}$) & & ($\times 10^{-1}$) & & \\ \hline
  JLA & $-1.42 \pm 0.84$ & $0.22 \pm 0.47$ & --- & --- & $1.25 \pm 0.06$ & $2.63 \pm 0.07$& $-5.32 \pm 0.45$ & $-19.11 \pm 0.02 $& $3.26 \pm 0.27$& $0.86$ \\ 
    Union2.1 & $5.77 \pm 0.92$ & $-3.11 \pm 1.16$ & $14 \pm 4$ & $-70 \pm 15$ & $1.04 \pm 0.07$ & $2.30 \pm 0.06$ & $-3.20 \pm 1.66$ & $-19.51 \pm 0.01 $& $3.03 \pm 0.24$ & $0.84$ \\\botrule
\end{tabular} \label{tabchialfa}}
\end{table}

In  Table \ref{tabchialfa}  we show  the results of the statistical analyses   described before for  the case where the luminosity distance is modified through $\alpha$ variation.   A  quick  look  at  this  table  reveals  that  the  Union~2.1
compilation seems to favor a dipole model, since the value of $\chi^2$
is smaller than  the one obtained with a  standard $\Lambda_{\rm CDM}$
cosmological  model.   For  the   JLA  compilation,  even  though  the
phenomenological dipole model results in  a lower value of the reduced
$\chi^2$,  the differences  between  the values  of  $\chi^2$ are  not
statistically significant. Thus,  we judge that this  dataset does not
yield significant evidence favoring the phenomenological dipole model.

\begin{figure*}[pb]
\centering
    \includegraphics[width=0.4\textwidth,trim=0.1cm 0.1cm 0.1cm 0.1cm,clip]{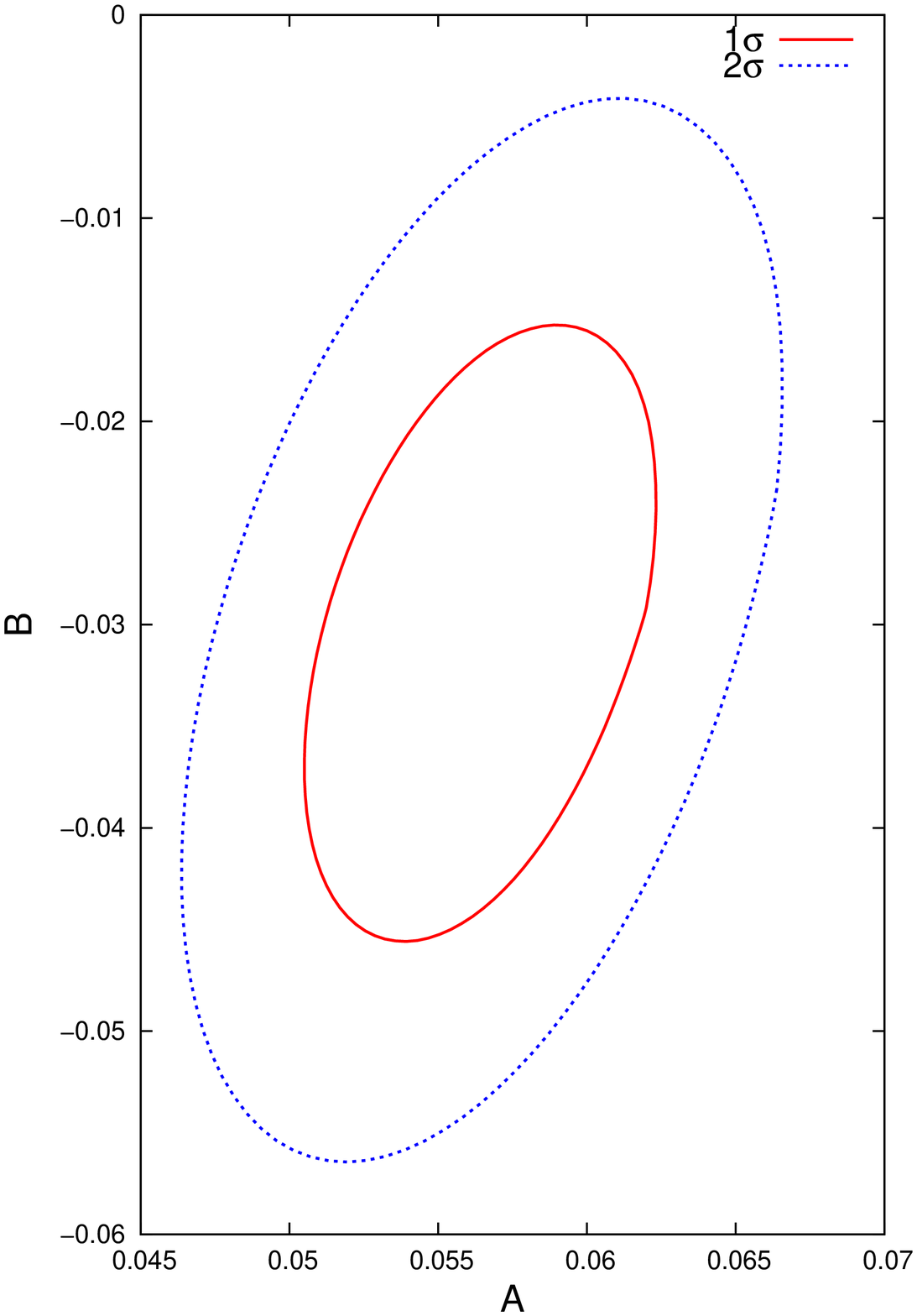}
    \includegraphics[width=0.4\textwidth,trim=0.1cm 0.1cm 0.1cm 0.1cm,clip]{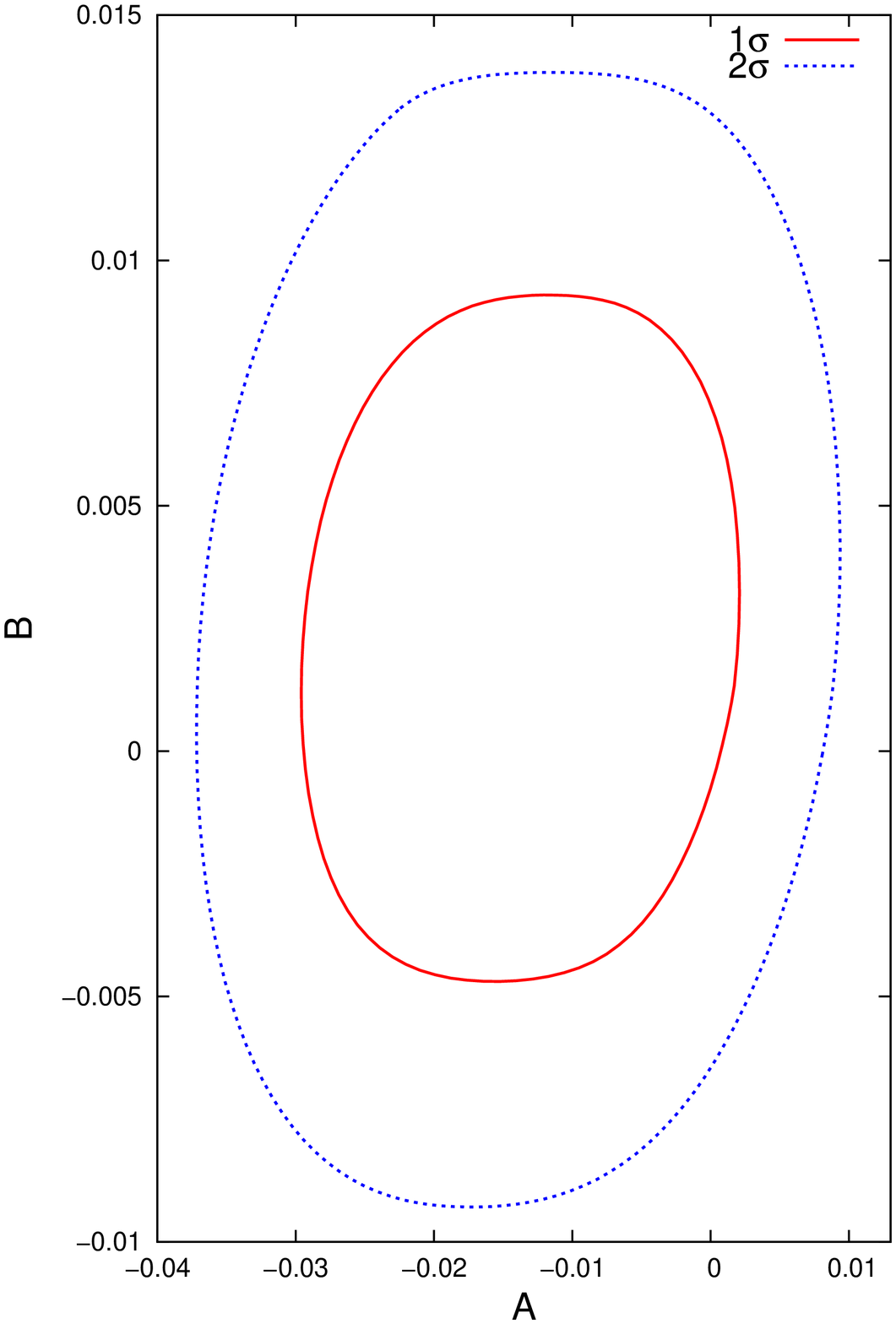}
    \vspace*{8pt}
 \caption{Results  of  the  statistical analysis  for  a  hypothetical
   variation of $\alpha$. We show the $1\sigma$ ($68\%$) and $2\sigma$
   ($95\%$) confidence  contours for the free  parameters $A_{\alpha}$
   and  $B_{\alpha}$ of  the  phenomenological  dipole model  obtained
   using the  Union~2.1 data  set (left  panel) and  the JLA  data set
   (right panel).}
\label{fig:contornos}
\end{figure*}

It should  be noted  as well that  the JLA dataset  does not  allow to
derive the direction of the dipole.   The reason for this turns out to
be that  the magnitude  of the dipole  $B_{\alpha}$ is  small compared
with the uncertainties. In fact, the results for the confidence limits
on the  dipole term $B_{\alpha}$  is consistent  with $0$ for  the JLA
analysis. Consequently, there is no difference between the theoretical
distance moduli  calculated for different dipole  directions.  This is
not the case for the Union~2.1 analysis and therefore, the statistical
analysis performed with this dataset yields limits on the direction of
the  dipole,  although  with  sizeable uncertainties  ($75\%$  in  right
ascension and  $12\%$ in declination). We  also note that there  is no
degeneracy between the free  parameters $A_{\alpha}$ and $B_{\alpha}$, for the analysis performed with the JLA data set while there is a little degeneracy for the Union2.1 case (see Fig.~\ref{fig:contornos}).

Table \ref{tabchialfa} also shows the values of the SNe Ia nuissance parameters and the cosmological density matter from the statistical analysis considering $\alpha$ as a varying constant for both compilations. For Union 2.1, the estimates of $\widehat{\delta}$ and $\Omega_{\rm m}$ are consistent at 1$\sigma$ level with those obtained by Ref. \refcite{Suzuki12}, while $\widehat{\alpha}$ and $\widehat{\beta}$ show a 2$\sigma$ level consistency. Besides, the ${\rm M}_{\rm b}$ estimates are only compatible at 5$\sigma$ level with those of Ref. \refcite{Suzuki12}. On the other hand, for JLA compilation, the estimates of two parameters $\Omega_{\rm m}$ and $\Delta_{\rm M}$ are in agreement with the ones from Ref. \refcite{Betoule14} at 1$\sigma$ level, $\widehat{\alpha}$ and ${\rm M}_{\rm b}$ values are consistent at 2$\sigma$ level, and only the $\widehat{\beta}$ estimates are consistent at 4$\sigma$ level. It should be mentioned that only the $\Omega_{\rm m}$ estimates of Refs. \refcite{Suzuki12} and \refcite{Betoule14} are consistent with each other at 1$\sigma$ level.

\begin{figure*}[pb]
\begin{center}
  \includegraphics[width=0.32\textwidth,trim=0cm 4.2cm 0.5cm 4cm,clip]{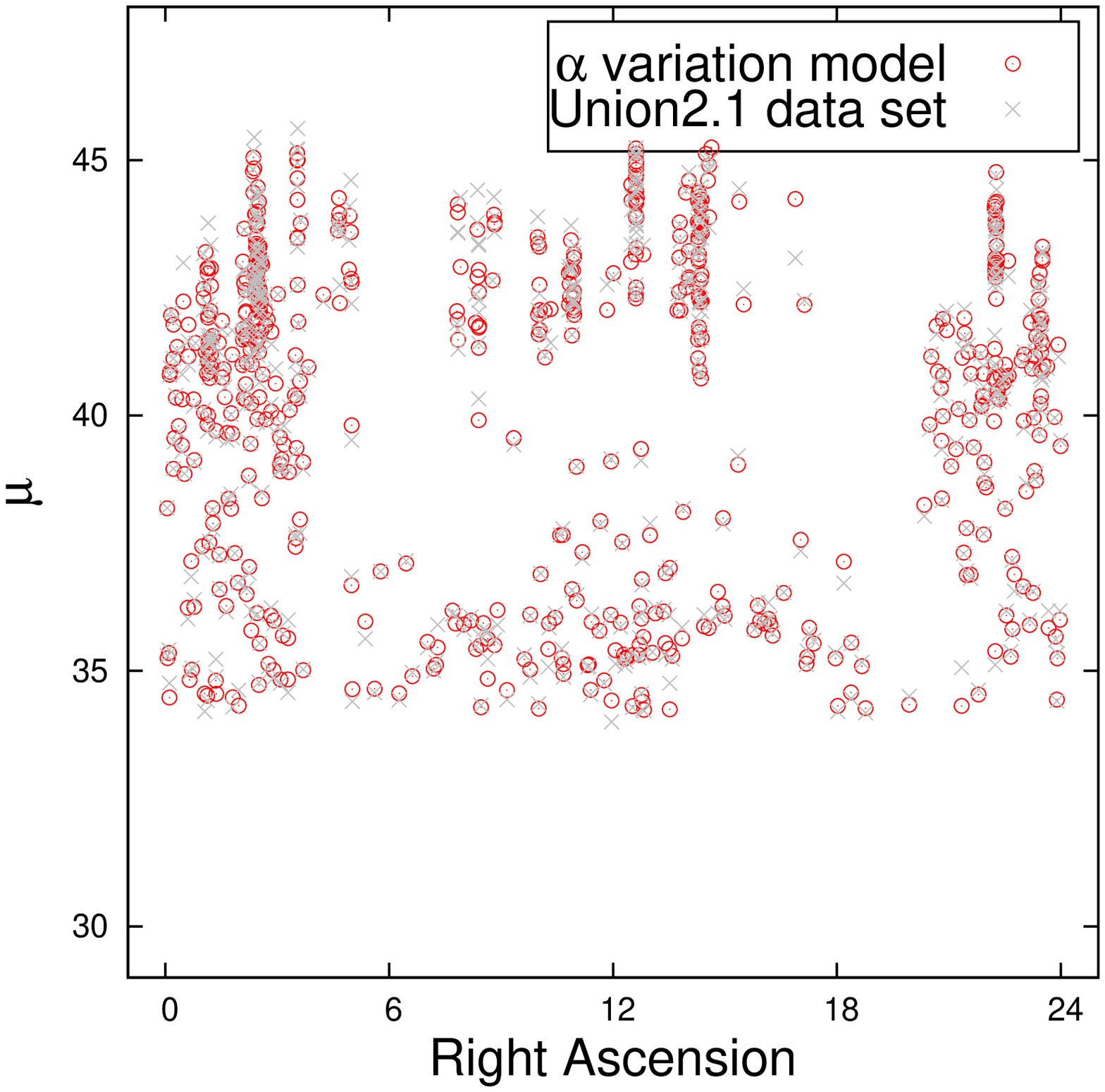}
  \includegraphics[width=0.32\textwidth,trim=0cm 4.2cm 0.5cm 4cm,clip]{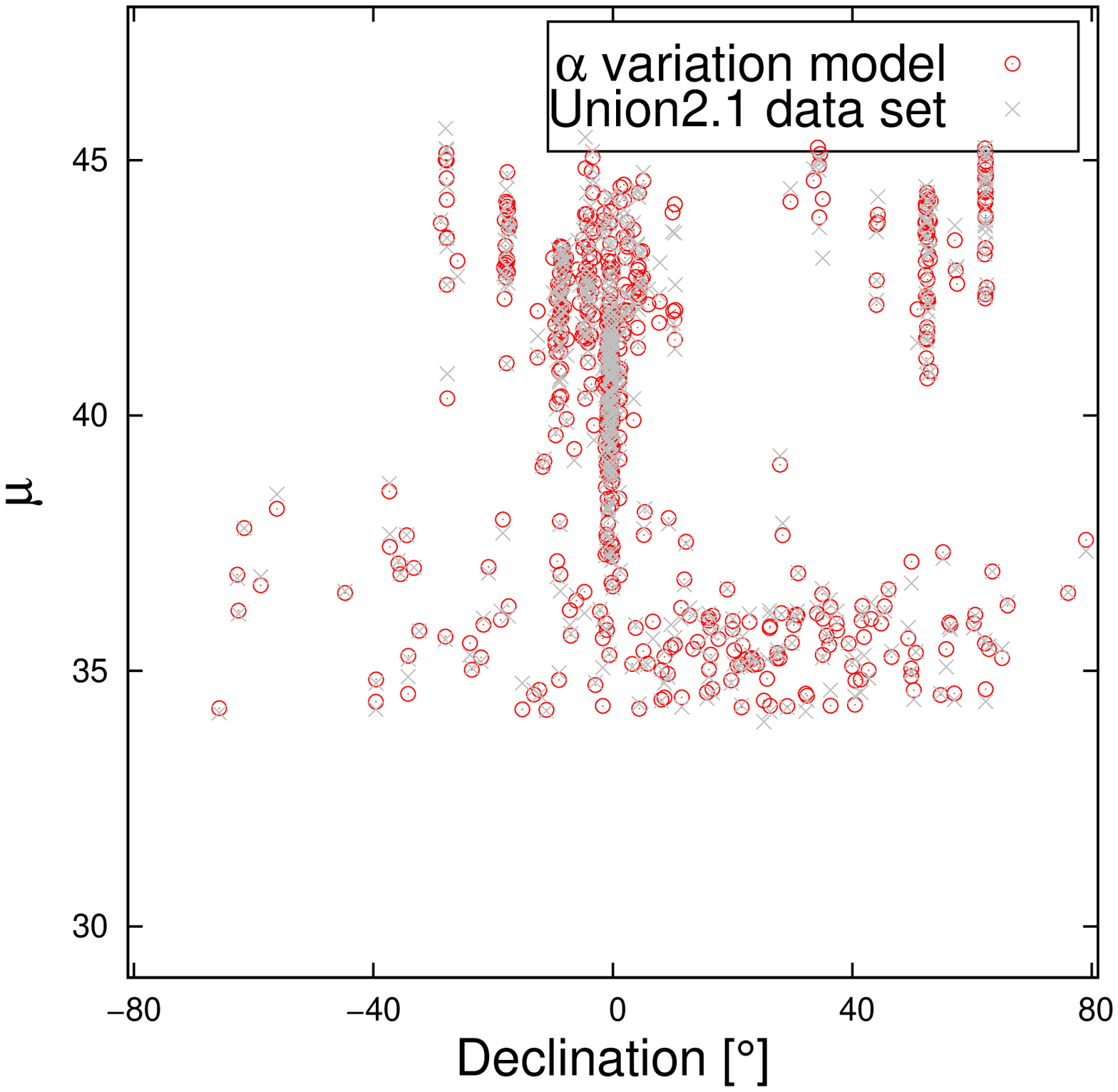}
  \includegraphics[width=0.32\textwidth,trim=0cm 4.2cm 0.5cm 4cm,clip]{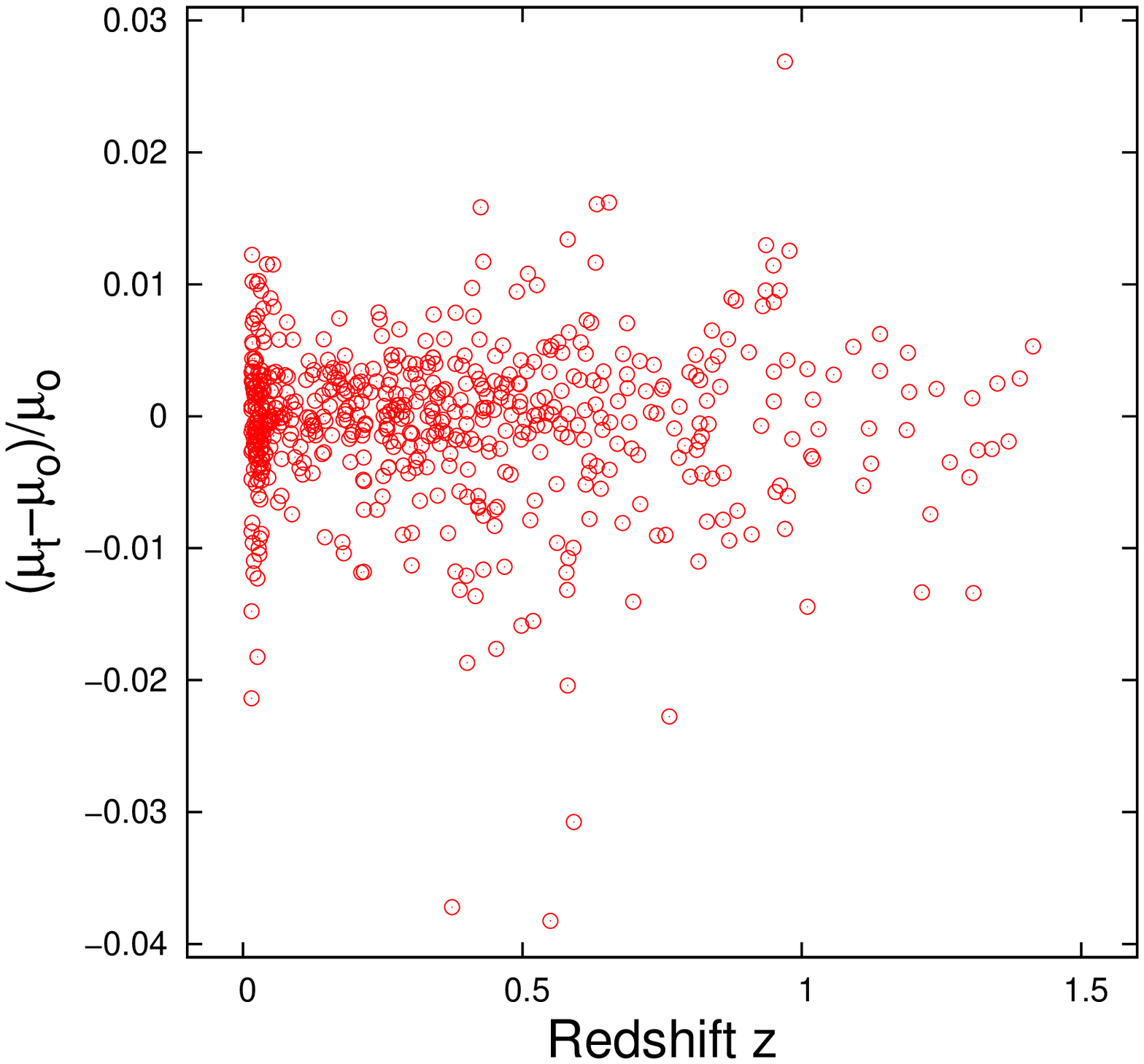}
  \includegraphics[width=0.32\textwidth,trim=0cm 4.2cm 0.5cm 4cm,clip]{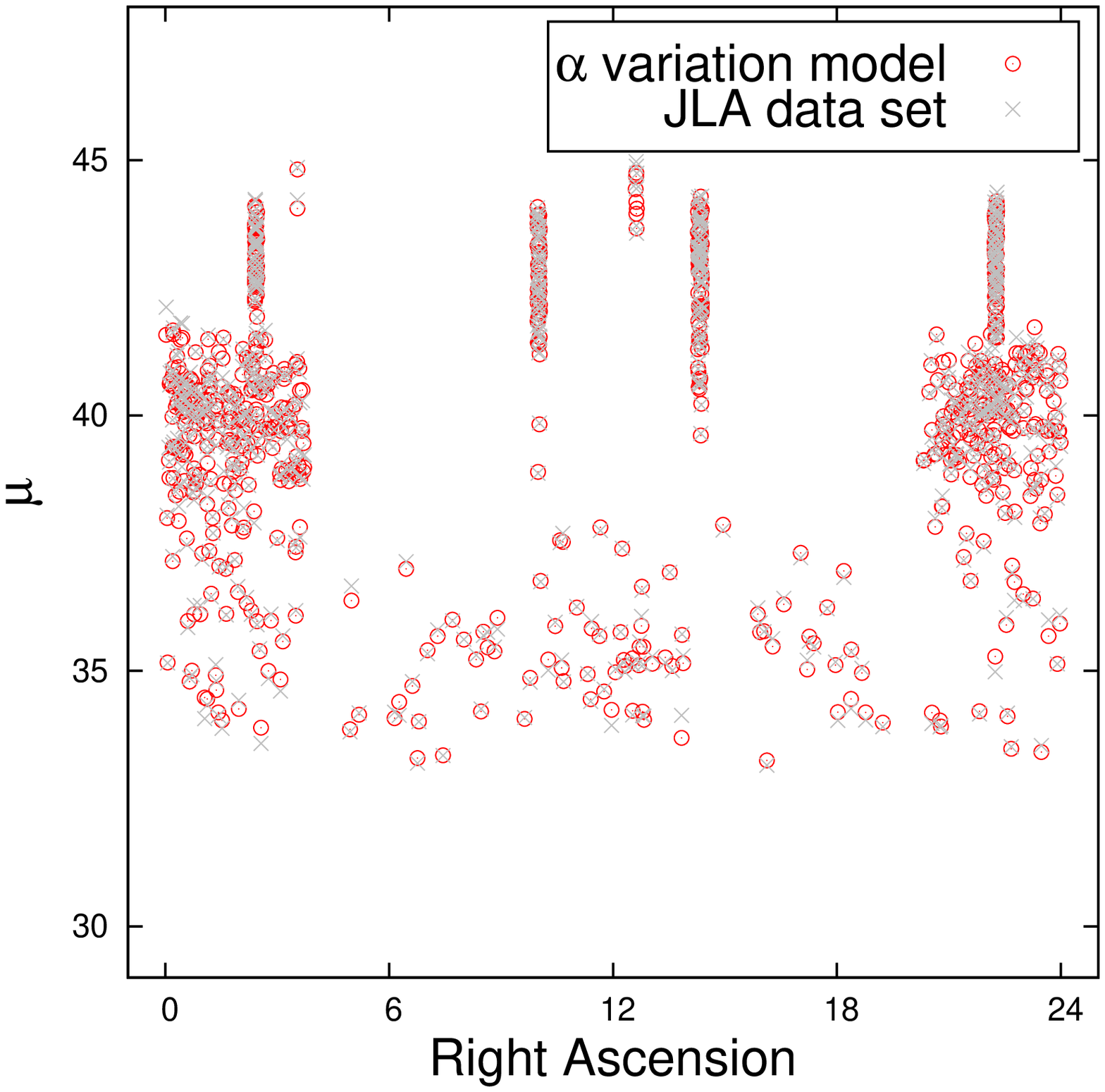}
  \includegraphics[width=0.32\textwidth,trim=0cm 4.2cm 0.5cm 4cm,clip]{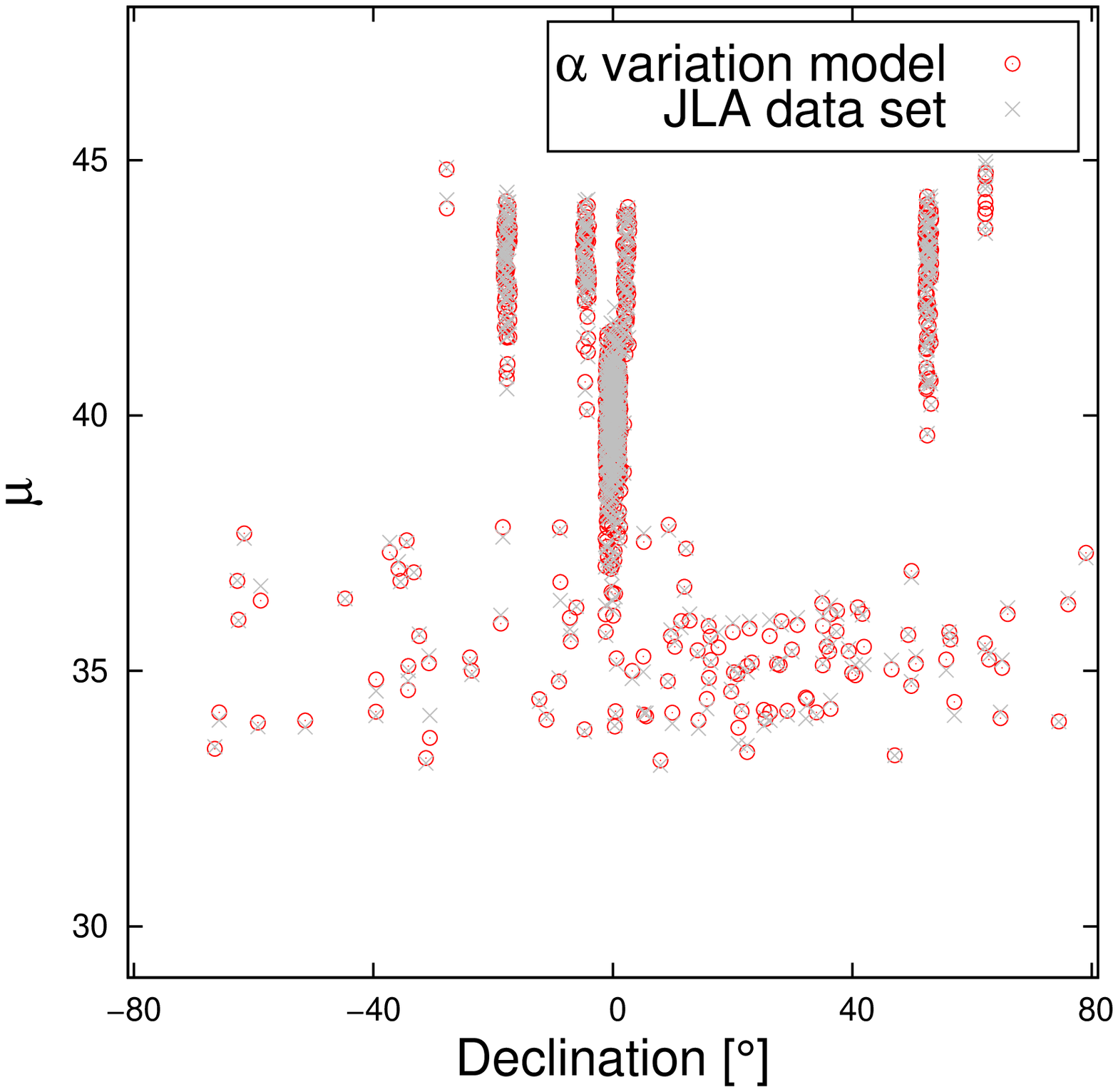}
  \includegraphics[width=0.32\textwidth,trim=0cm 4.2cm 0.5cm 4cm,clip]{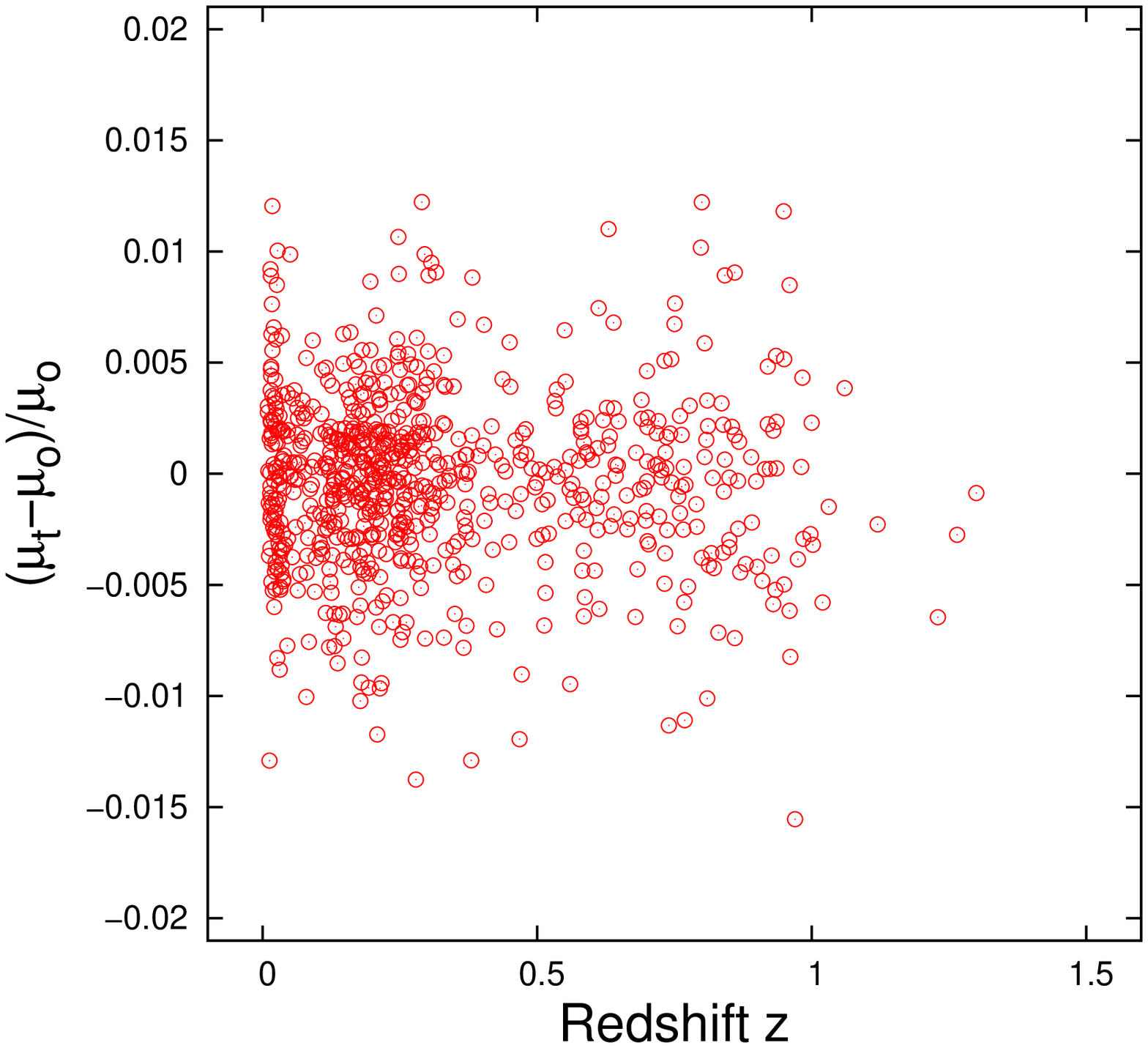}
  \vspace*{8pt}
\caption{Distance modulus as  a function of the  right ascension (left
  panels) and declination (middle panels) for the 
  Union~2.1 (upper panels) and JLA (lower panels) datasets. We compare
  the predictions of the phenomenological dipole model for a variation
  of $\alpha$ (red circles) with the observed data (gray symbols). In the right panels 
  we show the relative differences between the  predictions and the observed data as a function
  of the redshift.}
\label{fig:modulos}
\end{center}
\end{figure*}

Finally, in Fig.~\ref{fig:modulos} we  compare the individual distance
moduli obtained using the phenomenological model for $\alpha$ variation with the observed data using the SNe Ia nuissance parameters and $\Omega_{\rm m}$ estimates from the statistical analysis explained above.
The  upper panels  correspond to  the Union~2.1  dataset, whereas  the
bottom plots  correspond to  the JLA dataset.   Also, the  left panels
display  the distance  moduli as  a function  of right  ascension, the
central panels  show the individual  distance moduli as a  function of
declination and,  finally, in  the right panels  we plot the relative differences between the  predictions and the observed data as a function
  of the redshift. The   grey points are  the observed
data,  while  the  red  ones  are the  results  of  our  theoretical
calculations.

We continue  our analysis  examining a possible  variation of  $c$. In
Table~\ref{tab:modelo2}  we present  the  results  of the  statistical
analysis considering $c$  as a free parameter and  using the Union~2.1
and the  JLA datasets  respectively. As  in the  case for  the spatial
variation  of $\alpha$,  the  value of  $\chi^2/\nu$  does not  differ
significantly from  the value obtained  using the standard  model when
the JLA dataset is used.  However, when the Union~2.1 dataset is used,
the differences between the dipole model and the standard cosmological
model is sizeable. As it occurs for the case of a varying $\alpha$, the
JLA sample does  not allow to establish a preferred  direction for the
dipole. The reason  for this is the same previously  discussed for the
case of  a varying $\alpha$. It  is important to emphasize  that these
results set an independent upper bound to a possible spatial variation
of $c$. To  the best of our  knowledge this is the  first constraint on
such hypothetical variation.

\begin{table}
\tbl{Same as Table~\ref{tabchialfa}, but  for a model in which
     a variation of $c$ is adopted.}  
{\begin{tabular}{@{}cccccccccccc@{}}\toprule
 Dataset & $A_{\rm c}$  & $B_{\rm c}$ &  R.A. & $\delta$ & $\widetilde{\alpha}$ &  $\widetilde{\beta}$ & $\Delta$ & ${\rm M}_{\rm b}$& $\Omega_{\rm m}$& $\chi^2_{\nu}$ \\  
 & ($\times 10^{-1}$) & ($\times 10^{-2}$) & (h) & ($^{\circ}$)  & ($\times 10^{-1}$) & & ($\times 10^{-2}$) & & ($\times 10^{-1}$) & & \\ \hline 
JLA & $-2.49 \pm 0.09$  & $0 \pm 0.4$ & --- & --- & $1.26 \pm 0.06$ & $2.63 \pm 0.07$ & $-5.29 \pm 2.56$ & $-20.22 \pm 0.03$ & $3.26 \pm 0.47$ &  $0.88$ \\ 
  Union2.1 & $1.9 \pm 0.05$ & $1.47 \pm 0.6$ & $14 \pm 5$ & $-70 \pm 14$ & $1.04 \pm 0.07$ & $2.30 \pm 0.06$& $-3.25 \pm 2.10$ & $-18.61 \pm 0.02 $& $3.01 \pm 0.25$& $0.84$ \\\botrule
\end{tabular} \label{tab:modelo2}}
\end{table}

   In addition, we add  the SNe Ia nuissance parameters and the cosmological density matter estimates from the statistical analysis considering varying $c$ for both compilations in Table \ref{tab:modelo2}. For Union 2.1, the values of $\widehat{\delta}$ and $\Omega_{\rm m}$ are compatible  with those from Ref. \refcite{Suzuki12} at 1$\sigma$ level, while $\widehat{\alpha}$ and $\widehat{\beta}$ show a 2$\sigma$ level consistency. Finally, the estimates of ${\rm M}_{\rm b}$ are not consistent at all. Besides, for JLA compilation, the estimates of $\Delta_{\rm M}$ and $\Omega_{\rm m}$ are in agreement with the ones from Ref. \refcite{Betoule14} at 1$\sigma$ level while the constraints on  $\widehat{\alpha}$   are consistent at 2$\sigma$ level, and only the $\widehat{\beta}$ estimates are consistent at 4$\sigma$ level. Finally, there is no agreement between our  estimates on $M_b$ and those of Ref. \refcite{Betoule14} within 5 $\sigma$. However, it is importan to stress, that the Union2.1 and JLA results were obtained assuming a standard cosmological model which is not the case for the analyses performed in this paper. Most important, the estimates on $M_b$ provided by  Cepheid based calibrations of the SN Ia peak
luminosity (and therefore independent of the assumed cosmological model)  \cite{Saha99} are consistent within 5 $\sigma$ with the estimates of Table \ref{tab:modelo2}.

\section{Summary and Conclusions}
\label{sec:discusion}

In this paper we have analyzed  the dependence of the distance modulus
of Type Ia  supernovae on the fundamental constants $\alpha$ and $c$. In our  analysis we have included the dependence on
a possible variation of these constants  of the energy released in the
explosion. This study  had  not been  performed  before. Besides the dependence of the SNe I standardization with the distance modulus predicted by the theoretical model here exposed, we have also had the possibility of  estimating $\Omega_{\rm m}$ and the nuisance parameters of SNe Ia.  Using  the
scaling  laws  resulting   from  this  study  we   have  examined  the
possibility of obtaining  upper bounds to a  hypothetical variation of
these   fundamental  constants.    To   do  this   we   have  used   a
phenomenological  model that  accounts  for a  possible anisotropy  of
cosmological observables.   Specifically, we  have assumed  that these
fundamental constants, as suggested  by some recent observations, have
a  dipolar  dependence.   We  then compared  the  predictions  of  our
theoretical  analysis   with  the  most  up-to-date   compilations  of
observational data for thermonuclear  supernovae, namely the Union~2.1
and JLA  datasets.  The reason  to choose  Type Ia supernovae  for our
analysis is  threefold. On one  hand, the peak luminosities  of SNe~Ia
depend on $\alpha$ and $c$. Therefore, a  variation of these
constants  directly  translates  into   a  different  peak  bolometric
magnitude.   This,  in  turn,  means  that  the  distance  modulus  is
modified. On  the other hand, normal SNe~Ia  are a very homogeneous  class of
objects   that  can   be  observed   up  to   very  large   distances.
Consequently,  any possible  variation  of  the fundamental  constants
analyzed in this work would become prominent. Finally, the last reason
to adopt Type Ia supernovae for our analysis is that we have databases
of  observational measurements for a large number of them.

Our results  show that  the JLA  compilation of  SNe~Ia data  favors a
standard  cosmological  model,  in  which none  of  these  fundamental
constants  varies. On  the  contrary, when  the  Union~2.1 dataset  is
employed there is marginal evidence for such variations. Specifically,
the JLA dataset does not allow to obtain a preferred dipole direction.
This is not  the case when the Union~2.1 compilation  is used. However
evidence for  such variation is weak  at the $3\sigma$ level,  even in
the  case  in  which  the  best data  of  the  Union~2.1  database  is
employed. Thus, we  conclude that at $3\sigma$, the  parameters of the
supernova data are consistent with a null variation of the fundamental
constants.   This can  be viewed  from a  different perspective.   The
analysis of the observational data can  be used to set upper limits to
the spatial variation of the  fundamental constants.  We
obtained upper  bounds to the  spatial variation of $\alpha$ and $c$.  These   upper  limits   are  $\delta\alpha/\alpha\sim 10^{-2}$ and  $\delta  c/c  \sim  10^{-1}$.
To  the  best  of  our
knowledge, here we have reported the first upper limit to a hypothetical spatial variation of $c$.  Hence, we
judge that this is perhaps the  main result of our calculations. 
  On the other hand,  the best current upper limit on the spatial variation of $\alpha$ are $\delta \alpha/\alpha \sim 10^{-5}$, and was obtained using data from quasar absorption systems for a range of redshifts $0.3 < z < 2.8$ \cite{King12,Murphy17}. Clearly, this limit is considerably more stringent that that obtained here. However, we emphasize that the upper bound reported here has been obtained using a totally independent method and a different dataset (the range of redshifts are  $0.623 < z < 1.415$ for the Union2.1 and $0.05 < z < 1$ for JLA), and therefore complements  (and is compatible with) the previous upper limit.

\section*{Acknowledgments}

C.   Negrelli, L.   Kraiselburd and  S.  Landau  are supported  by PIP
11220120100504     (CONICET),     grant     G140     (UNLP),     while
E. Garc\'{i}a--Berro is supported by MINECO grant AYA2014-59084-P, and
by the AGAUR.

\appendix

\bibliographystyle{ws-ijmpd}
\bibliography{testIJMPD}

\begin{thebibliography}{10}

\bibitem{Pablo}
P.~{Lor{\'e}n-Aguilar}, E.~{Garc{\'{\i}}a-Berro}, J.~{Isern} and Y.~A.
  {Kubyshin}, {\em Classical Quant. Grav.} {\bf 20}  (2003) 3885,
  \href{http://arxiv.org/abs/astro-ph/0309722}{{\ttfamily astro-ph/0309722}}.

\bibitem{Uzan}
J.-P. {Uzan}, {\em Living Reviews in Relativity} {\bf 14}  (2011)  ~2,
  \href{http://arxiv.org/abs/1009.5514}{{\ttfamily arXiv:1009.5514
  [astro-ph.CO]}}.

\bibitem{Yuri}
E.~{Garc{\'{\i}}a-Berro}, J.~{Isern} and Y.~A. {Kubyshin}, {\em \aapr} {\bf 14}
   (2007) 113.

\bibitem{bekenstein82}
J.~D. {Bekenstein}, {\em \prd} {\bf 25}  (1982) 1527.

\bibitem{bekenstein2002}
J.~D. {Bekenstein}, {\em \prd} {\bf 66}  (2002)   123514,
  \href{http://arxiv.org/abs/gr-qc/0208081}{{\ttfamily gr-qc/0208081}}.

\bibitem{bsm02}
J.~D. {Barrow}, H.~B. {Sandvik} and J.~{Magueijo}, {\em \prd} {\bf 65}  (2002)
   063504, \href{http://arxiv.org/abs/astro-ph/0109414}{{\ttfamily
  astro-ph/0109414}}.

\bibitem{moffat2}
J.~W. {Moffat}, {\em Int. J. Mod. Phys. D} {\bf 2}  (1993) 351,
  \href{http://arxiv.org/abs/gr-qc/9211020}{{\ttfamily gr-qc/9211020}}.

\bibitem{albrecht}
A.~{Albrecht} and J.~{Magueijo}, {\em \prd} {\bf 59}  (1999)   043516,
  \href{http://arxiv.org/abs/astro-ph/9811018}{{\ttfamily astro-ph/9811018}}.

\bibitem{barrow}
J.~D. {Barrow}, {\em \prd} {\bf 59}  (1999)   043515,
  \href{http://arxiv.org/abs/astro-ph/9811022}{{\ttfamily astro-ph/9811022}}.

\bibitem{BM05}
J.~D. {Barrow} and J.~{Magueijo}, {\em \prd} {\bf 72}  (2005)   043521,
  \href{http://arxiv.org/abs/astro-ph/0503222}{{\ttfamily astro-ph/0503222}}.

\bibitem{dam3}
T.~{Damour} and F.~{Dyson}, {\em Nuclear Physics B} {\bf 480}  (1996) 37,
  \href{http://arxiv.org/abs/hep-ph/9606486}{{\ttfamily hep-ph/9606486}}.

\bibitem{Petrov06}
Y.~V. {Petrov}, A.~I. {Nazarov}, M.~S. {Onegin}, V.~Y. {Petrov} and E.~G.
  {Sakhnovsky}, {\em \prc} {\bf 74}  (2006)   064610,
  \href{http://arxiv.org/abs/hep-ph/0506186}{{\ttfamily hep-ph/0506186}}.

\bibitem{Gould06}
C.~R. {Gould}, E.~I. {Sharapov} and S.~K. {Lamoreaux}, {\em \prc} {\bf 74}
  (2006)   024607, \href{http://arxiv.org/abs/nucl-ex/0701019}{{\ttfamily
  nucl-ex/0701019}}.

\bibitem{dyson}
F.~J. Dyson, {\em Phys. Rev. Lett.} {\bf 19}  (1967) 1291.

\bibitem{sisterna}
P.~Sisterna and H.~Vucetich, {\em Phys. Rev. D} {\bf 41}  (1990) 1034.

\bibitem{Olive04b}
K.~A. {Olive}, M.~{Pospelov}, Y.-Z. {Qian}, G.~{Manh{\`e}s},
  E.~{Vangioni-Flam}, A.~{Coc} and M.~{Cass{\'e}}, {\em \prd} {\bf 69}  (2004)
   027701, \href{http://arxiv.org/abs/astro-ph/0309252}{{\ttfamily
  astro-ph/0309252}}.

\bibitem{pres}
J.~D. {Prestage}, R.~L. {Tjoelker} and L.~{Maleki}, {\em Phys. Rev. Lett.} {\bf
  74}  (1995) 3511.

\bibitem{Peik04}
E.~{Peik}, B.~{Lipphardt}, H.~{Schnatz}, T.~{Schneider}, C.~{Tamm} and S.~G.
  {Karshenboim}, {\em Phys. Rev. Lett.} {\bf 93}  (2004)   170801,
  \href{http://arxiv.org/abs/physics/0402132}{{\ttfamily physics/0402132}}.

\bibitem{rosenband08}
T.~{Rosenband}, D.~B. {Hume}, P.~O. {Schmidt}, C.~W. {Chou}, A.~{Brusch},
  L.~{Lorini}, W.~H. {Oskay}, R.~E. {Drullinger}, T.~M. {Fortier}, J.~E.
  {Stalnaker}, S.~A. {Diddams}, W.~C. {Swann}, N.~R. {Newbury}, W.~M. {Itano},
  D.~J. {Wineland} and J.~C. {Bergquist}, {\em Science} {\bf 319}  (2008) 1808.

\bibitem{bahc}
J.~N. {Bahcall}, C.~L. {Steinhardt} and D.~{Schlegel}, {\em \apj} {\bf 600}
  (2004) 520, \href{http://arxiv.org/abs/astro-ph/0301507}{{\ttfamily
  astro-ph/0301507}}.

\bibitem{lev}
S.~A. {Levshakov}, M.~{Dessauges-Zavadsky}, S.~{D'Odorico} and P.~{Molaro},
  {\em \mnras} {\bf 333}  (2002) 373,
  \href{http://arxiv.org/abs/astro-ph/0106194}{{\ttfamily astro-ph/0106194}}.

\bibitem{murphy1}
M.~T. {Murphy}, J.~K. {Webb}, V.~V. {Flambaum}, V.~A. {Dzuba}, C.~W.
  {Churchill}, J.~X. {Prochaska}, J.~D. {Barrow} and A.~M. {Wolfe}, {\em
  \mnras} {\bf 327}  (2001) 1208,
  \href{http://arxiv.org/abs/astro-ph/0012419}{{\ttfamily astro-ph/0012419}}.

\bibitem{murphy2}
M.~T. {Murphy}, J.~K. {Webb}, V.~V. {Flambaum}, J.~X. {Prochaska} and A.~M.
  {Wolfe}, {\em \mnras} {\bf 327}  (2001) 1237,
  \href{http://arxiv.org/abs/astro-ph/0012421}{{\ttfamily astro-ph/0012421}}.

\bibitem{Webb99}
J.~K. {Webb}, V.~V. {Flambaum}, C.~W. {Churchill}, M.~J. {Drinkwater} and J.~D.
  {Barrow}, {\em Phys. Rev. Lett.} {\bf 82}  (1999) 884,
  \href{http://arxiv.org/abs/astro-ph/9803165}{{\ttfamily astro-ph/9803165}}.

\bibitem{webb2}
J.~K. {Webb}, M.~T. {Murphy}, V.~V. {Flambaum}, V.~A. {Dzuba}, J.~D. {Barrow},
  C.~W. {Churchill}, J.~X. {Prochaska} and A.~M. {Wolfe}, {\em Phys. Rev.
  Lett.} {\bf 87}  (2001)   091301,
  \href{http://arxiv.org/abs/astro-ph/0012539}{{\ttfamily astro-ph/0012539}}.

\bibitem{Galli13}
S.~{Galli}, {\em \prd} {\bf 87}  (2013)   123516,
  \href{http://arxiv.org/abs/1212.1075}{{\ttfamily arXiv:1212.1075}}.

\bibitem{Holanda16}
R.~F.~L. {Holanda}, S.~J. {Landau}, J.~S. {Alcaniz}, I.~E. {S{\'a}nchez G.} and
  V.~C. {Busti}, {\em \jcap} {\bf 5}  (2016)   047,
  \href{http://arxiv.org/abs/1510.07240}{{\ttfamily arXiv:1510.07240}}.

\bibitem{Holanda16b}
R.~F.~L. {Holanda}, V.~C. {Busti}, L.~R. {Cola{\c c}o}, J.~S. {Alcaniz} and
  S.~J. {Landau}, {\em \jcap} {\bf 8}  (2016)   055,
  \href{http://arxiv.org/abs/1605.02578}{{\ttfamily arXiv:1605.02578}}.

\bibitem{Martino16}
I.~{de Martino}, C.~J.~A.~P. {Martins}, H.~{Ebeling} and D.~{Kocevski}, {\em
  \prd} {\bf 94}  (2016)   083008,
  \href{http://arxiv.org/abs/1605.03053}{{\ttfamily arXiv:1605.03053}}.

\bibitem{Martino16b}
I.~{de Martino}, C.~J.~A.~P. {Martins}, H.~{Ebeling} and D.~{Kocevski}, {\em
  Universe} {\bf 2}  (2016)  ~34,
  \href{http://arxiv.org/abs/1612.06739}{{\ttfamily arXiv:1612.06739}}.

\bibitem{bergstrom}
L.~{Bergstr{\"o}m}, S.~{Iguri} and H.~{Rubinstein}, {\em \prd} {\bf 60}  (1999)
    045005, \href{http://arxiv.org/abs/astro-ph/9902157}{{\ttfamily
  astro-ph/9902157}}.

\bibitem{mosquera}
M.~E. {Mosquera} and O.~{Civitarese}, {\em \aap} {\bf 551}  (2013)   A122.

\bibitem{bat}
R.~A. {Battye}, R.~{Crittenden} and J.~{Weller}, {\em \prd} {\bf 63}  (2001)
  043505, \href{http://arxiv.org/abs/astro-ph/0008265}{{\ttfamily
  astro-ph/0008265}}.

\bibitem{avelino}
P.~P. {Avelino}, C.~J.~A.~P. {Martins}, G.~{Rocha} and P.~{Viana}, {\em \prd}
  {\bf 62}  (2000)   123508,
  \href{http://arxiv.org/abs/astro-ph/0008446}{{\ttfamily astro-ph/0008446}}.

\bibitem{Planck2015}
{Planck Collaboration}, P.~A.~R. {Ade}, N.~{Aghanim}, M.~{Arnaud},
  M.~{Ashdown}, J.~{Aumont}, C.~{Baccigalupi}, A.~J. {Banday}, R.~B.
  {Barreiro}, E.~{Battaner}, K.~{Benabed}, A.~{Benoit-L{\'e}vy}, J.-P.
  {Bernard}, M.~{Bersanelli}, P.~{Bielewicz}, J.~R. {Bond}, J.~{Borrill}, F.~R.
  {Bouchet}, C.~{Burigana}, R.~C. {Butler}, E.~{Calabrese}, A.~{Chamballu},
  H.~C. {Chiang}, P.~R. {Christensen}, D.~L. {Clements}, L.~P.~L. {Colombo},
  F.~{Couchot}, A.~{Curto}, F.~{Cuttaia}, L.~{Danese}, R.~D. {Davies}, R.~J.
  {Davis}, P.~{de Bernardis}, A.~{de Rosa}, G.~{de Zotti}, J.~{Delabrouille},
  J.~M. {Diego}, H.~{Dole}, O.~{Dor{\'e}}, X.~{Dupac}, T.~A. {En{\ss}lin},
  H.~K. {Eriksen}, O.~{Fabre}, F.~{Finelli}, O.~{Forni}, M.~{Frailis},
  E.~{Franceschi}, S.~{Galeotta}, S.~{Galli}, K.~{Ganga}, M.~{Giard},
  J.~{Gonz{\'a}lez-Nuevo}, K.~M. {G{\'o}rski}, A.~{Gregorio}, A.~{Gruppuso},
  F.~K. {Hansen}, D.~{Hanson}, D.~L. {Harrison}, S.~{Henrot-Versill{\'e}},
  C.~{Hern{\'a}ndez-Monteagudo}, D.~{Herranz}, S.~R. {Hildebrandt}, E.~{Hivon},
  M.~{Hobson}, W.~A. {Holmes}, A.~{Hornstrup}, W.~{Hovest}, K.~M.
  {Huffenberger}, A.~H. {Jaffe}, W.~C. {Jones}, E.~{Keih{\"a}nen},
  R.~{Keskitalo}, R.~{Kneissl}, J.~{Knoche}, M.~{Kunz}, H.~{Kurki-Suonio},
  J.-M. {Lamarre}, A.~{Lasenby}, C.~R. {Lawrence}, R.~{Leonardi},
  J.~{Lesgourgues}, M.~{Liguori}, P.~B. {Lilje}, M.~{Linden-V{\o}rnle},
  M.~{L{\'o}pez-Caniego}, P.~M. {Lubin}, J.~F. {Mac{\'{\i}}as-P{\'e}rez},
  N.~{Mandolesi}, M.~{Maris}, P.~G. {Martin},
  E.~{Mart{\'{\i}}nez-Gonz{\'a}lez}, S.~{Masi}, S.~{Matarrese}, P.~{Mazzotta},
  P.~R. {Meinhold}, A.~{Melchiorri}, L.~{Mendes}, E.~{Menegoni}, A.~{Mennella},
  M.~{Migliaccio}, M.-A. {Miville-Desch{\^e}nes}, A.~{Moneti}, L.~{Montier},
  G.~{Morgante}, A.~{Moss}, D.~{Munshi}, J.~A. {Murphy}, P.~{Naselsky},
  F.~{Nati}, P.~{Natoli}, H.~U. {N{\o}rgaard-Nielsen}, F.~{Noviello},
  D.~{Novikov}, I.~{Novikov}, C.~A. {Oxborrow}, L.~{Pagano}, F.~{Pajot},
  D.~{Paoletti}, F.~{Pasian}, G.~{Patanchon}, O.~{Perdereau}, L.~{Perotto},
  F.~{Perrotta}, F.~{Piacentini}, M.~{Piat}, E.~{Pierpaoli}, D.~{Pietrobon},
  S.~{Plaszczynski}, E.~{Pointecouteau}, G.~{Polenta}, N.~{Ponthieu},
  L.~{Popa}, G.~W. {Pratt}, S.~{Prunet}, J.~P. {Rachen}, R.~{Rebolo},
  M.~{Reinecke}, M.~{Remazeilles}, C.~{Renault}, S.~{Ricciardi},
  I.~{Ristorcelli}, G.~{Rocha}, G.~{Roudier}, B.~{Rusholme}, M.~{Sandri},
  G.~{Savini}, D.~{Scott}, L.~D. {Spencer}, V.~{Stolyarov}, R.~{Sudiwala},
  D.~{Sutton}, A.-S. {Suur-Uski}, J.-F. {Sygnet}, J.~A. {Tauber},
  D.~{Tavagnacco}, L.~{Terenzi}, L.~{Toffolatti}, M.~{Tomasi}, M.~{Tristram},
  M.~{Tucci}, J.-P. {Uzan}, L.~{Valenziano}, J.~{Valiviita}, B.~{Van Tent},
  P.~{Vielva}, F.~{Villa}, L.~A. {Wade}, D.~{Yvon}, A.~{Zacchei} and
  A.~{Zonca}, {\em \aap} {\bf 580}  (2015)   A22,
  \href{http://arxiv.org/abs/1406.7482}{{\ttfamily arXiv:1406.7482}}.

\bibitem{obryan15}
J.~{O'Bryan}, J.~{Smidt}, F.~{De Bernardis} and A.~{Cooray}, {\em \apj} {\bf
  798}  (2015)  ~18.

\bibitem{Murphy03b}
M.~T. {Murphy}, J.~K. {Webb} and V.~V. {Flambaum}, {\em \mnras} {\bf 345}
  (2003) 609, \href{http://arxiv.org/abs/astro-ph/0306483}{{\ttfamily
  astro-ph/0306483}}.

\bibitem{Webb11}
J.~K. {Webb}, J.~A. {King}, M.~T. {Murphy}, V.~V. {Flambaum}, R.~F. {Carswell}
  and M.~B. {Bainbridge}, {\em Phys. Rev. Lett.} {\bf 107}  (2011)   191101,
  \href{http://arxiv.org/abs/1008.3907}{{\ttfamily arXiv:1008.3907
  [astro-ph.CO]}}.

\bibitem{King12}
J.~A. {King}, J.~K. {Webb}, M.~T. {Murphy}, V.~V. {Flambaum}, R.~F. {Carswell},
  M.~B. {Bainbridge}, M.~R. {Wilczynska} and F.~E. {Koch}, {\em \mnras} {\bf
  422}  (2012) 3370, \href{http://arxiv.org/abs/1202.4758}{{\ttfamily
  arXiv:1202.4758 [astro-ph.CO]}}.

\bibitem{whit}
J.~B. {Whitmore} and M.~T. {Murphy}, {\em \mnras} {\bf 447}  (2015) 446,
  \href{http://arxiv.org/abs/1409.4467}{{\ttfamily arXiv:1409.4467
  [astro-ph.IM]}}.

\bibitem{PM16}
A.~M.~M. {Pinho} and C.~J.~A.~P. {Martins}, {\em Phys. Lett. B} {\bf 756}
  (2016) 121, \href{http://arxiv.org/abs/1603.04498}{{\ttfamily
  arXiv:1603.04498}}.

\bibitem{Murphy16}
M.~T. {Murphy}, A.~L. {Malec} and J.~X. {Prochaska}, {\em \mnras} {\bf 461}
  (2016) 2461, \href{http://arxiv.org/abs/1606.06293}{{\ttfamily
  arXiv:1606.06293}}.

\bibitem{Murphy17}
M.~T. {Murphy}, A.~L. {Malec} and J.~X. {Prochaska}, {\em \mnras} {\bf 464}
  (2017) 2609.

\bibitem{Iorio11}
L.~{Iorio}, {\em \mnras} {\bf 417}  (2011) 2392,
  \href{http://arxiv.org/abs/1104.5192}{{\ttfamily arXiv:1104.5192 [gr-qc]}}.

\bibitem{Li15}
X.~{Li}, H.-N. {Lin}, S.~{Wang} and Z.~{Chang}, {\em European Physical Journal
  C} {\bf 75}  (2015)   181, \href{http://arxiv.org/abs/1501.06738}{{\ttfamily
  arXiv:1501.06738 [gr-qc]}}.

\bibitem{YWC14}
X.~{Yang}, F.~Y. {Wang} and Z.~{Chu}, {\em \mnras} {\bf 437}  (2014) 1840.

\bibitem{Mariano}
A.~{Mariano} and L.~{Perivolaropoulos}, {\em \prd} {\bf 86}  (2012)   083517,
  \href{http://arxiv.org/abs/1206.4055}{{\ttfamily arXiv:1206.4055
  [astro-ph.CO]}}.

\bibitem{Mariano13}
A.~{Mariano} and L.~{Perivolaropoulos}, {\em \prd} {\bf 87}  (2013)   043511,
  \href{http://arxiv.org/abs/1211.5915}{{\ttfamily arXiv:1211.5915
  [astro-ph.CO]}}.

\bibitem{Chang15}
Z.~{Chang} and H.-N. {Lin}, {\em \mnras} {\bf 446}  (2015) 2952,
  \href{http://arxiv.org/abs/1411.1466}{{\ttfamily arXiv:1411.1466}}.

\bibitem{Lin16b}
H.-N. {Lin}, X.~{Li} and Z.~{Chang}, {\em \mnras} {\bf 460}  (2016) 617,
  \href{http://arxiv.org/abs/1604.07505}{{\ttfamily arXiv:1604.07505}}.

\bibitem{Lin16}
H.-N. {Lin}, S.~{Wang}, Z.~{Chang} and X.~{Li}, {\em \mnras} {\bf 456}  (2016)
  1881, \href{http://arxiv.org/abs/1504.03428}{{\ttfamily arXiv:1504.03428}}.

\bibitem{Betoule14}
M.~{Betoule}, R.~{Kessler}, J.~{Guy}, J.~{Mosher}, D.~{Hardin}, R.~{Biswas},
  P.~{Astier}, P.~{El-Hage}, M.~{Konig}, S.~{Kuhlmann}, J.~{Marriner},
  R.~{Pain}, N.~{Regnault}, C.~{Balland}, B.~A. {Bassett}, P.~J. {Brown},
  H.~{Campbell}, R.~G. {Carlberg}, F.~{Cellier-Holzem}, D.~{Cinabro},
  A.~{Conley}, C.~B. {D'Andrea}, D.~L. {DePoy}, M.~{Doi}, R.~S. {Ellis},
  S.~{Fabbro}, A.~V. {Filippenko}, R.~J. {Foley}, J.~A. {Frieman},
  D.~{Fouchez}, L.~{Galbany}, A.~{Goobar}, R.~R. {Gupta}, G.~J. {Hill},
  R.~{Hlozek}, C.~J. {Hogan}, I.~M. {Hook}, D.~A. {Howell}, S.~W. {Jha}, L.~{Le
  Guillou}, G.~{Leloudas}, C.~{Lidman}, J.~L. {Marshall}, A.~{M{\"o}ller},
  A.~M. {Mour{\~a}o}, J.~{Neveu}, R.~{Nichol}, M.~D. {Olmstead},
  N.~{Palanque-Delabrouille}, S.~{Perlmutter}, J.~L. {Prieto}, C.~J.
  {Pritchet}, M.~{Richmond}, A.~G. {Riess}, V.~{Ruhlmann-Kleider}, M.~{Sako},
  K.~{Schahmaneche}, D.~P. {Schneider}, M.~{Smith}, J.~{Sollerman},
  M.~{Sullivan}, N.~A. {Walton} and C.~J. {Wheeler}, {\em \aap} {\bf 568}
  (2014)   A22, \href{http://arxiv.org/abs/1401.4064}{{\ttfamily
  arXiv:1401.4064}}.

\bibitem{quasarspol}
D.~{Hutsem{\'e}kers}, L.~{Braibant}, V.~{Pelgrims} and D.~{Sluse}, {\em \aap}
  {\bf 572}  (2014)   A18, \href{http://arxiv.org/abs/1409.6098}{{\ttfamily
  arXiv:1409.6098}}.

\bibitem{anomaliesCMB}
C.~J. {Copi}, D.~{Huterer}, D.~J. {Schwarz} and G.~D. {Starkman}, {\em Adv. in
  Astron.} {\bf 2010}  (2010)   847541.

\bibitem{krai}
L.~{Kraiselburd}, S.~J. {Landau}, C.~{Negrelli} and E.~{Garc{\'{\i}}a-Berro},
  {\em \apss} {\bf 357}  (2015)  ~4,
  \href{http://arxiv.org/abs/1412.3418}{{\ttfamily arXiv:1412.3418}}.

\bibitem{ChibaKohri}
T.~{Chiba} and K.~{Kohri}, {\em Progress of Theoretical Physics} {\bf 110}
  (2003) 195.

\bibitem{MBS02}
J.~{Magueijo}, J.~D. {Barrow} and H.~B. {Sandvik}, {\em Phys. Lett. B} {\bf
  549}  (2002) 284, \href{http://arxiv.org/abs/astro-ph/0202374}{{\ttfamily
  astro-ph/0202374}}.

\bibitem{Magueijo01}
J.~a. Magueijo, {\em Phys. Rev. D} {\bf 63}  (2001)   043502.

\bibitem{Suzuki12}
N.~{Suzuki}, D.~{Rubin}, C.~{Lidman}, G.~{Aldering}, R.~{Amanullah},
  K.~{Barbary}, L.~F. {Barrientos}, J.~{Botyanszki}, M.~{Brodwin},
  N.~{Connolly}, K.~S. {Dawson}, A.~{Dey}, M.~{Doi}, M.~{Donahue},
  S.~{Deustua}, P.~{Eisenhardt}, E.~{Ellingson}, L.~{Faccioli}, V.~{Fadeyev},
  H.~K. {Fakhouri}, A.~S. {Fruchter}, D.~G. {Gilbank}, M.~D. {Gladders},
  G.~{Goldhaber}, A.~H. {Gonzalez}, A.~{Goobar}, A.~{Gude}, T.~{Hattori},
  H.~{Hoekstra}, E.~{Hsiao}, X.~{Huang}, Y.~{Ihara}, M.~J. {Jee},
  D.~{Johnston}, N.~{Kashikawa}, B.~{Koester}, K.~{Konishi}, M.~{Kowalski},
  E.~V. {Linder}, L.~{Lubin}, J.~{Melbourne}, J.~{Meyers}, T.~{Morokuma},
  F.~{Munshi}, C.~{Mullis}, T.~{Oda}, N.~{Panagia}, S.~{Perlmutter},
  M.~{Postman}, T.~{Pritchard}, J.~{Rhodes}, P.~{Ripoche}, P.~{Rosati}, D.~J.
  {Schlegel}, A.~{Spadafora}, S.~A. {Stanford}, V.~{Stanishev}, D.~{Stern},
  M.~{Strovink}, N.~{Takanashi}, K.~{Tokita}, M.~{Wagner}, L.~{Wang},
  N.~{Yasuda}, H.~K.~C. {Yee} and T.~{Supernova Cosmology Project}, {\em \apj}
  {\bf 746}  (2012)  ~85, \href{http://arxiv.org/abs/1105.3470}{{\ttfamily
  arXiv:1105.3470 [astro-ph.CO]}}.

\bibitem{riess}
A.~G. {Riess}, A.~V. {Filippenko}, W.~{Li}, R.~R. {Treffers}, B.~P. {Schmidt},
  Y.~{Qiu}, J.~{Hu}, M.~{Armstrong}, C.~{Faranda}, E.~{Thouvenot} and
  C.~{Buil}, {\em \aj} {\bf 118}  (1999) 2675,
  \href{http://arxiv.org/abs/astro-ph/9907037}{{\ttfamily astro-ph/9907037}}.

\bibitem{Karp77}
A.~H. {Karp}, G.~{Lasher}, K.~L. {Chan} and E.~E. {Salpeter}, {\em \apj} {\bf
  214}  (1977) 161.

\bibitem{Planck16}
{Planck Collaboration}, P.~A.~R. {Ade}, N.~{Aghanim}, M.~{Arnaud},
  M.~{Ashdown}, J.~{Aumont}, C.~{Baccigalupi}, A.~J. {Banday}, R.~B.
  {Barreiro}, J.~G. {Bartlett} and et~al., {\em \aap} {\bf 594}  (2016)   A13,
  \href{http://arxiv.org/abs/1502.01589}{{\ttfamily arXiv:1502.01589}}.

\bibitem{Beutler11}
F.~{Beutler}, C.~{Blake}, M.~{Colless}, D.~H. {Jones}, L.~{Staveley-Smith},
  L.~{Campbell}, Q.~{Parker}, W.~{Saunders} and F.~{Watson}, {\em \mnras} {\bf
  416}  (2011) 3017, \href{http://arxiv.org/abs/1106.3366}{{\ttfamily
  arXiv:1106.3366}}.

\bibitem{BAO1}
L.~Anderson {\em et~al.}, {\em \mnras} {\bf 441}  (2014) 24.

\bibitem{Ross15}
A.~J. {Ross}, L.~{Samushia}, C.~{Howlett}, W.~J. {Percival}, A.~{Burden} and
  M.~{Manera}, {\em \mnras} {\bf 449}  (2015) 835,
  \href{http://arxiv.org/abs/1409.3242}{{\ttfamily arXiv:1409.3242}}.

\bibitem{Phillips}
M.~M. {Phillips}, {\em \apjl} {\bf 413}  (1993) L105.

\bibitem{Planck14}
{Planck Collaboration}, P.~A.~R. {Ade}, N.~{Aghanim}, C.~{Armitage-Caplan},
  M.~{Arnaud}, M.~{Ashdown}, F.~{Atrio-Barandela}, J.~{Aumont},
  C.~{Baccigalupi}, A.~J. {Banday} and et~al., {\em \aap} {\bf 571}  (2014)
  A16, \href{http://arxiv.org/abs/1303.5076}{{\ttfamily arXiv:1303.5076}}.

\bibitem{Saha99}
A.~{Saha}, A.~{Sandage}, G.~A. {Tammann}, L.~{Labhardt}, F.~D. {Macchetto} and
  N.~{Panagia}, {\em \apj} {\bf 522} (September 1999) 802,
  \href{http://arxiv.org/abs/astro-ph/9904389}{{\ttfamily astro-ph/9904389}}.

\end{thebibliography}


\end{document}